\documentclass{article}
\usepackage{graphicx} 
\usepackage[top=1in, bottom=0.9in, left=0.9in, right=1in]{geometry}
\usepackage[textwidth=8em,textsize=small]{todonotes}
\usepackage[utf8]{inputenc}
\usepackage{mathbbol}
\usepackage{bbm}
\usepackage{bm}
\usepackage{url}
\usepackage{amsmath}
\usepackage{natbib}
\usepackage[utf8]{inputenc}
\usepackage{xcolor}
\usepackage{graphicx}
\usepackage{enumerate}
\usepackage{amsthm}
\usepackage{amsmath}
\usepackage{amssymb}
\usepackage{marginnote}
\usepackage{mathtools}
\usepackage{float}

\usepackage{booktabs}
\usepackage{changepage}
\usepackage{ulem}
\usepackage{color}
\usepackage{setspace}
\usepackage{lscape}
\usepackage{algorithm}
\usepackage{algpseudocode}
\newtheorem{theorem}{Theorem}[section]

\begin{document}
\begin{center}
{\Large Coalescent Inference for Epidemics with Latent Periods
}
\\
Isaac H. Goldstein$^1$, 
Julia Palacios$^1$ 
\\
$^1$Department of Statistics, Stanford University
\end{center}

\begin{abstract}
  Coalescent models are used to study the transmission dynamics of
  rapidly evolving pathogens from molecular sequence data obtained
  from infected individuals.
  However coalescent parameters, such as effective population size,
  offer limited interpretability for transmission dynamics.
  In this work, we derive a coalescent model for exposed–infected
  population dynamics that allows us to infer the number of infected
  individuals and the effective reproduction number over time from
  the sample genealogy.
  The model can be interpreted as a two-deme model in which
  coalescence is restricted to individuals from different demes
  (exposed and infected).
  We propose a new data-augmentation framework with Phase-type
  distribution for Bayesian inference of epidemiological parameters.
  We study the performance of our approach on simulations and apply
  it to re-analyze the 2014 Ebola outbreak in Liberia.
\end{abstract}

\section{Introduction}
It is now common practice to collect pathogen molecular sequence data obtained from infected individuals during an infectious disease outbreak or epidemic \citep{attwood2022phylogenetic, cappello2022statistical}. 
These sequences can be used to infer the shared transmission history of the pathogen among sampled
individuals, represented by a genealogy, by modeling a process of mutations
along the branches of the genealogy
\citep{salemi2009phylogenetic,didier2024models}. Typically, the
genealogy is assumed to \textit{a priori} depend on the population
dynamics through two widely used classes of models: coalescent models
and birth-death-sampling (BDS) models \citep{tavare2004lectures,
stadler2010sampling}.

The coalescent \citep{kingman1982coalescent}, initially proposed as a
scaled limiting process under Wright-Fisher population dynamics, has
been shown to be a good approximation to the distribution of the
genealogy of a random sample of individuals under a broad class of
discrete time population models \citep{mohle2000ancestral}. The
coalescent model is typically parametrized in terms of the effective
population size which corresponds to the population size under
Wright-Fisher dynamics. However, interpretation of this parameter is
more challenging under more general population dynamics. In
\cite{volz2009phylodynamics} and \cite{volz2012complex}, the authors
proposed structured coalescent models under mechanistic deterministic
epidemic population dynamics, including birth-death-migration models
described by ordinary differential equations (ODEs), explicitly
linking the effective population size to the parameters and states of
the epidemic population process. Most notably, the derivation of the
coalescent model in this setting does not rely on time-scaling,
limiting approximations or small sampling proportions.

The birth-death-sampling model (BDS), as in the structured
coalescent, also assumes the population process is a general
birth-death stochastic process. However, this approach incorporates
an explicit sampling model that allows sampling in bulk with
pre-specified probabilities or sampling through time according to a
point process. This is one key difference with the
structured coalescent that instead conditions on the number of
samples at given sampling times. The BDS implementations estimate
birth, death, and sampling rates, and it has been extended for
multi-type processes  \citep{kuhnert2016phylodynamics}. In
particular, the multi-type birth-death process model was recently
developed to estimate epidemic population sizes in addition to
population parameters  \citep{vaughan2025bayesian}. A review of some
of these methods is available in \cite{cappello2022statistical}.
Recently, \cite{king2025exact}  proposed a framework that assumes
general Markov population processes and derives a genealogical model
that allows likelihood computation in a forward time fashion.

In this manuscript, we assume a birth-death-migration population
process in which individuals experience a delay between being
infected (exposed, E-compartment) and becoming infectious
(infectious, I-compartment), a common characteristic of many real
world pathogens \citep{Xin2021,velasquez2015time}.
This simplified EI population model allows us to avoid modeling the
number of susceptibles. Moreover, its deterministic form governed by
differential equations, has a closed form solution which is fast to
compute \citep{goldstein2024semiparametric}, and that we leverage in our inferential framework.
We derive the corresponding (structured) coalescent process as a
thinned point process under certain assumptions.
Our derivation is most similar to \citep{volz2012complex}, however we
start from stochastic population dynamics forward in time and couple
it with the coalescent backwards in time conditioned on observed
samples at given sampling times.
As in \cite{volz2012complex}, our derivation does not rely on
scaling, limiting approximations or small sampling proportions.

Multiple approaches for Bayesian inference of population parameters
under the aforementioned models have been proposed in recent years,
including Markov chain Monte Carlo (MCMC), and sequential Monte Carlo
approaches.
The methods developed in
\cite{volz2012complex,muller2017structured,volz2018bayesian}, infer
population parameters under a wide range of deterministic structured
population processes by integrating over the states of sampled lineages.
This is made tractable by approximating the likelihood assuming
lineage states are independent \citep{muller2017structured,volz2018bayesian}.
For the simplest epidemic process, the SIR stochastic model, which
assumes a single class of infectious individuals with no latent
period, \cite{rasmussen2011inference} 
used particle MCMC to infer the latent number of infectious individuals.
More recently, \cite{tang2023fitting} used a linear noise
approximation, an approximate continuous stochastic process, to model
the latent number of infectious individuals 
in a more computationally efficient manner.

In this manuscript, we develop a Bayesian data augmentation method to
estimate model parameters and population size trajectories from a
genealogy. We target the posterior distribution of model parameters,
population size trajectories, and compartment states of sampled
lineages at observed times (coalescent and sampling times). For the
augmented likelihood we recast the coalescent density as phase type
distributions \citep{zeng2021studying,hobolth2024phase}.
Compartment states of sampled lineages at observed times are
sampled exactly, while population parameters and population size
trajectories are sampled using MCMC.
In our implementation, we assume the reproduction number is a
piece-wise constant trajectory modeled
with a Gaussian Markov random field prior.
We test our new method on simulated epidemics and apply our model to
the 2014 Ebola outbreak in Liberia.
\section{Methods}
\subsection{Phylogeny as Data}
Throughout this paper, we will assume a fixed genealogy as the ``observed data". 
This genealogy is normally estimated from molecular sequences, or from previous studies.
The relevant characteristics of the genealogy for our model are the times of coalescence and the times when lineages are sampled, as well as the specific type of event at those times. We represent the times of events with the vector $\mathbf{t}=\{t_{0}, \dots, t_{n}\}$, where $n$ is the total number of coalescent and sampling times, and $t_{0} < \dots < t_{n}$. 
Further, these times are measured in units backwards in time, so that $t_{0} = 0$ is the present time, and $t_{N}$ is the time of the most recent common ancestor of the sampled lineages (Figure \ref{fig:Fig1}(C)).
Let the vector $\mathbf{x} = \{x_{0}, \dots, x_{n}\}$ be the types of these events, where if $x_{i} = 1$, the event at time $t_{i}$ is a coalescent event and if $x_{i} = 0$ the event at time $t_{i}$ is a sampling event.
\subsection{The EI Population Model}
The EI population model assumes an epidemic evolving in forward time, with the population split into two compartments, individuals who are infected but not infectious (E), and individuals who are infectious (I). 
To avoid confusion, we will use the letter $u$ to represent forward time, and the letter $t$ for backwards time.
The EI model is a bivariate
continuous time Markov jump process 
$\mathbf{H}(u) = (E(u), I(u))$, that satisfies:
\begin{align*}
    \text{P}(\mathbf{H}(u + du) &= (e + 1, i)  \mid \mathbf{H}(u) = (e,i)) = \alpha(u)i du + o(du),\\
    \text{P}(\mathbf{H}(u + du) &= (e - 1, i + 1) \mid \mathbf{H}(u) = (e,i)) =  \gamma e du + o(du), \\ 
    \text{P}(\mathbf{H}(u + du) &= (e, i - 1) \mid \mathbf{H}(u) = (e,i)) = \nu  i  du + o(du)\\
    \text{P}(\mathbf{H}(u + du) &= (e, i) \mid \mathbf{H}(u) = (e,i)) = 1- (\alpha(u) i + \gamma e  + \nu i) du + o(du).
\end{align*}
Here $\gamma$ is interpreted as the inverse of the mean latent period, $\nu$ is the inverse of the mean infectious period, and $\alpha(u)$ is the time-varying per-capita number of new infections per unit time. Throughout we will assume $\alpha(u)$ is a continuous function. 
In keeping with terminology from prior phylogenetics literature, we will refer to the event when $\mathbf{H}(u + du) = (e + 1, i)$ as a birth, and the event when $\mathbf{H}(u + du) = (e - 1, i + 1)$ as a migration. 
The effective reproduction number (which is often denoted $R_{t}$, but for consistency we will use $R_{u}$), is then defined as $R_{u} = \alpha(u)/\nu$. 
Appendix Section \ref{app:comp_models} describes in more detail the differences between the EI population model and the more traditional SEIR population model. 

\begin{figure}[H]
  \includegraphics[width=\textwidth]{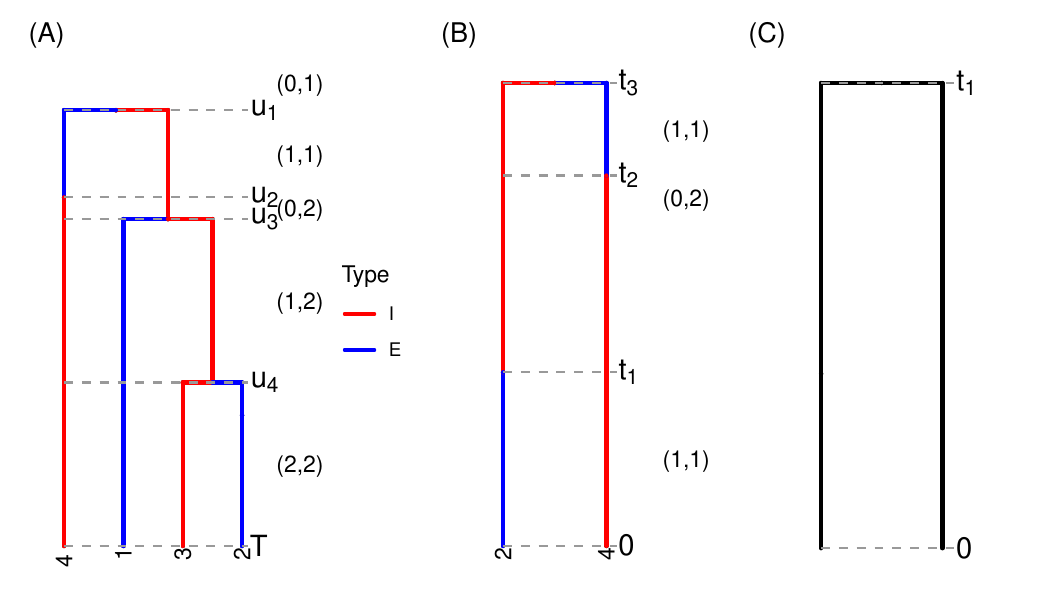}
  \caption{(A) A realization of the EI population process with
    initial state $\mathbf{H}(0)=(0,1)$ and end state
    $\mathbf{H}(T)=(2,2)$, together with one corresponding population genealogy. Given $C(T)=(n_{E}(T),n_{I}(T))=(1,1)$, one
    E lineage (blue) and one I lineage (red) are sampled at time $T$.
    Figure (B) shows the corresponding EI sampled process (and sample genealogy) when tips 4
    and 2 are chosen. Here time increases from tips to root, with
    $t_{3}=T$. In (C), only coalescent events (births) of sampled
  lineages are observed.}
  \label{fig:Fig1}
\end{figure}

Although the $\mathbf{H}(u)$ process only keeps track of the number of individuals in each compartment, we can represent the population history by an EI population genealogy in which the event $\mathbf{H}(u + du) = (e + 1, i)$ corresponds to a birth when a uniformly sampled I lineage (red) splits into two lineages: one E (blue) lineage and one I (red) lineage in the population genealogy; and the event  $\mathbf{H}(u + du) = (e - 1, i + 1)$ corresponds to a migration when a uniformly sampled E (blue) lineage becomes an I (red) lineage; and the event $\mathbf{H}(u + du) = (e , i - 1)$ corresponds to a death when a uniformly sampled I lineage is pruned in the population genealogy. Figure \ref{fig:Fig1}(A) shows a realization of the
process from state H$(0)=(0,1)$ to H$(T)=(2,2)$, and a
corresponding population genealogy with lineages colored according to their
state (exposed or infectious). For illustration purposes only, tree
tips are labeled. We note that even when tips are not labeled in the population genealogy, many tree shapes correspond to the same $\mathbf{H}(u)$ trajectory (for example lineage 4 could have been chosen to split at time $u_{4}$ giving rise to a different tree shape). For our likelihood expressions we focus on $\mathbf{H}(u)$ realizations. The corresponding population genealogy likelihood would include a combinatorial factor that accounts for the number of different tree shapes. Similarly, the corresponding population labeled genealogy likelihood would include a combinatorial factor that accounts for the number of different labeled tree topologies.

\subsection{The EI sampled 
genealogy}

Assume a population that evolves (forward in time) according to the
EI population model started at $\mathbf{H}(0)=(0,1)$. Given the full
realization $(\mathbf{H}(u))_{u\in [0,T]}$ until present day at time
$T$, we sample uniformly at random $n_{E}$ from the total $E(T)$ of
exposed individuals at time $T$, and $n_{I}$ from the total $I(T)$ of
infected individuals. We then trace back the number of sampled individuals in each compartment over time to obtain the sample EI continuous time process
$\mathbf{C}(u)=(n_{E}(u),n_{I}(u))$. 
Figure~\ref{fig:Fig1}(B) shows the genealogy of samples 2
and 4 obtained by sampling $n_{E}=1$ and $n_{I}=1$ exposed and
infected lineages at time $T$ from the realization of
Figure~\ref{fig:Fig1}(A). For illustration purposes in
Figure~\ref{fig:Fig1}, we have labeled tips $1$ to $4$ and fixed a particular tree shape, however we
note that our $\mathbf{H}(u)$ process only counts the number of E and I individuals over time.

In this section we first derive the exact likelihood
  of the joint process of $\mathbf{H}(u)$, $\mathbf{C}(u)$, then an
  approximate likelihood of $\mathbf{H}(u), \mathbf{C}(u)$ when the
  population counts are approximated by a piecewise constant function, and finally
  we propose a backwards time process
  $\mathbf{\tilde{C}}(t)$, whose likelihood is the thinned
  likelihood of the joint approximated process.
  The likelihood of the backwards time process is what we
will use for inference.

The sample's transition history
$(\mathbf{C}(u))_{u\in[0,T]}$ is referred to as the fully observed EI
coalescent trajectory. $(\mathbf{C}(u))_{u\in[0,T]}$ can also be
constructed sequentially by first generating $(\mathbf{H}(u))_{u\in
[0,T]}$ and then sampling the discrete (lazy) jump chain
$\{\mathbf{C}(u_{i})\}_{i=k}^{1}$ from tips to root, assuming an
initial state $\mathbf{C}(T)=(n_{E}(T),n_{I}(T))$. We will show how to do this
through the following examples.

\textbf{Example 1a.} Assume we are given the EI realization shown in
Figure~\ref{fig:Fig1}(A) and that $n_{E}(T)=1$ and $n_{I}(T)=1$, that
is $\mathbf{C}(T)=(1,1)$. Then, given that a birth occurs at $u_{4}$,
this event is observed in the sample as a coalescent event with
probability $P(\mathbf{C}(u_{4})=(0,1) \mid
\mathbf{C}(T)=(1,1),(\mathbf{H}(u))_{u \in [0,T]})=1/4$, since we can
pick the E to be in the sample out of the two (w.p. $1/2$) and similarly for I.
Alternatively, $u_{4}$ could be observed in the sample as a migration
event if only the E lineage involved in $u_{4}$ is in the sample
(lineage 2 in Figure \ref{fig:Fig1}(A)).
This event also has probability $P(\mathbf{C}(u_{4})=(0,2) \mid
\mathbf{C}(T)=(1,1),(\mathbf{H}(u))_{u \in [0,T]})=1/4$.
This possibility is shown in Figure \ref{fig:Fig1}(B), where lineage
2 changes from blue to red at time $t_{1}$ in backwards time which is
equal to time $u_{4}$ in forwards time.

The probability that $u_{3}$ is in the sample as a coalescent event
is the probability that the E lineage involved is in the sample (
w.p. $1/2$) and the I lineage involved in $u_{4}$ is in the sample
(w.p 1/2), therefore $u_{3}$ is in the sample with probability
$P(\mathbf{C}(u_{3})=(0,1) \mid
\mathbf{C}(T)=(1,1),(\mathbf{H}(u))_{u \in [0,T]})=1/4$. \\
Similarly $u_{2}$ is in the sample (as migration) with probability
$P(\mathbf{C}(u_{2})=(1,1) \mid
\mathbf{C}(T)=(1,1),(\mathbf{H}(u))_{u \in [0,T]})=1/2$, and finally,
$u_{1}$ is in the sample (as coalescent) with probability
$P(\mathbf{C}(u_{1})=(0,1) \mid
\mathbf{C}(T)=(1,1),(\mathbf{H}(u))_{u \in [0,T]})=1/2$. 

\textbf{Example 1b.}
We will now compute the same probability for $u_{3}$ in a coalescent
manner (sequentially from tips to root). Conditioned on
$\mathbf{C}(T)=(1,1)$, $\mathbf{C}(u_{4})$ can take the values
$(1,1), (0,2),$ and $(0,1)$ with probabilities $1/2, 1/4$ and $1/4$
respectively. However $u_{3}$ cannot be a coalescent in the sample if
$\mathbf{C}(u_{4})$ takes the values $(0,2),(0,1)$. Therefore,
conditioned on $\mathbf{C}(u_{4})=(1,1)$, $u_{3}$ is a coalescent in
the sample with probability  $1/2$, and
$\text{P}(\mathbf{C}(u_{3})=(0,1)\mid
\mathbf{C}(T)=(1,1),\mathbf{H}(u))=\text{P}(\mathbf{C}(u_{3})=(0,1)\mid
\mathbf{C}(u_{4})=(1,1),\mathbf{H}(u)) \times
\text{P}(\mathbf{C}(u_{4})=(1,1)\mid \mathbf{C}(T)=(1,1),\mathbf{H}(u))=1/4$.

In practice, we assume that we only sample infectious individuals and
so $\mathbf{C}(T)=(0,n_{I}(T))$.
Modeling $\mathbf{H}(u)$ as a Marked point process, let $m(u) \in
\{1,2,3\}$ mark whether the transition type at time $u$ is a birth,
migration, or death. Then, the $\{\mathbf{C}(u_{i})\}$ jump chain is
governed by the following transition probabilities.\\

\noindent Coalescence of sampled E and I lineages:
\begin{align} \label{eq:trans1}
    \text{P}(\mathbf{C}(u_{i})=\mathbf{C}(u_{i+1})+(-1,0) \mid \mathbf{C}(u_{i+1}), m(u_{i})=1,(\mathbf{H}(t))_{t\in [0,T]})&=\frac{n_{E}(u_{i+1})}{E(u_{i})}\frac{n_{I}(u_{i+1})}{I(u_{i})}
    \end{align}
Given $(\mathbf{H}(u))_{u\in [0,T]}$ and a birth transition at time $u_{i}$ ($m(u_{i})=1$), we proceed backwards in time choosing two lineages uniformly at random to coalesce. Given there are $n_{E}(u_{i+1})$ sampled E lineages out of the total of $E(u^{-}_{i+1})=E(u_{i})$, and $n_{I}(u_{i+1})$ I sampled lineages out of the total of $I(u^{-}_{i+1})=I(u_{i})$, this coalescence will involve two lineages in the sample with probability given in Eq.~\ref{eq:trans1}. Recall that according to section 2.2, $I(t)$ and $E(t)$ are both right continuous functions.

In this case, the new state in the $\mathbf{C}(u)$ chain becomes $\mathbf{C}(u_{i})=(n_{E}(u_{i+1})-1,n_{I}(u_{i+1}))$. One E and one I lineage merge to become one I lineage and so the number of I lineages in the sample does not change.\\

\noindent Coalescence of sampled E and non-sampled I lineages:    
    \begin{align}
    \label{eq:trans2}
    \text{P}(\mathbf{C}(u_{i})=\mathbf{C}(u_{i+1})+(-1,1) \mid \mathbf{C}(u_{i+1}), m(u_{i})=1,(\mathbf{H}(t))_{t\in [0,T]})&=\frac{n_{E}(u_{i+1})}{E(u_{i})}\frac{I(u_{i})-n_{I}(u_{i+1})}{I(u_{i})}.
    \end{align}
Same conditioning as before, but coalescence occurs between an E lineage in the sample and an I lineage not in the sample. In this case, the E lineage in the sample becomes an I lineage after merging with I, and so $\mathbf{C}(u_{i})=(n_{E}(u_{i+1})-1,n_{I}(u_{i+1})+1)$. We refer to this event as coalescent migration.\\

\noindent Coalescence of non-sampled E and non-sampled I lineages or coalescence of non-sampled E and sampled I:   \begin{align}\label{eq:trans3}
     \text{P}(\mathbf{C}(u_{i})=\mathbf{C}(u_{i+1}) \mid \mathbf{C}(u_{i+1}), m(u_{i})=1,(\mathbf{H}(t))_{t\in [0,T]})&=1-\frac{n_{E}(u_{i+1})}{E(u_{i})}. 
    \end{align}
Same conditioning as before, but coalescence occurs between two lineages that are not in the sample or between an I lineage in the sample and an E lineage not in the sample. In both cases, the state of $\mathbf{C}(u_{i})$ does not change.  \\

\noindent Migration in the sample: \begin{align}\label{eq:trans4} \text{P}(\mathbf{C}(u_{i})=\mathbf{C}(u_{i+1})+(1,-1) \mid \mathbf{C}(u_{i+1}), m(u_{i})=2,(\mathbf{H}(t))_{t\in [0,T]})&=\frac{n_{I}(u_{i+1})}{I(u_{i})}.
    \end{align}

Given $(\mathbf{H}(u))_{u \in [0,T]}$, $\mathbf{C}(u_{i+1})$, and a migration transition at time $u_{i}$ ($m(u_{i})=2$), one I lineage in the sample is selected out of all $I(u_{i+1})$ to become an E lineage (in the sample). In this case, the state becomes $\mathbf{C}(u_{i})=(n_{E}(u_{i+1})+1,n_{I}(u_{i+1})-1)$.\\

Migration not in the sample:
    \begin{align} \label{eq:trans5}
     \text{P}(\mathbf{C}(u_{i})=\mathbf{C}(u_{i+1})\mid \mathbf{C}(u_{i+1}), m(u_{i})=2,(\mathbf{H}(t))_{t\in [0,T]})&=1-\frac{n_{I}(u_{i+1})}{I(u_{i})}.
\end{align}   
Same conditioning as before, but migration occurs in a lineage not in the sample.\\

We can now derive the likelihood of the joint process: the EI population process $\mathbf{H}(u)$ (forward in time), and the fully observed EI coalescent jump process $\{\mathbf{C}(u_{i})\}$ (proceeding backwards from time $T$ to 0), conditioned on $\mathbf{H}(0)=(0,1)$ and $\mathbf{C}(T)=(0,n_{I}(T))$. 
We note that $\mathbf{C}(T)$ imposes constraints in the EI population process as this implies $I(T)\geq n_{I}(T)$. 
In particular, having $n_{I}(T)>0$ samples implies that the EI population process did not go extinct in $[0,T]$. 
Thus, the likelihood of the EI population process $(\mathbf{H}(u))_{u \in [0,T]}$ conditioned on $\mathbf{H}(0)=(0,1)$ and $I(T) \geq n_{I}(T)$ is:

\begin{align}\label{eq:pop_cond}
 \mathcal{L}^{*}(\lambda, \nu, \alpha; \mathbf{H}(u))=e^{-\int^{T}_{u_{k}}\lambda(u)du}\prod^{k}_{i=1}f(u_{i}\mid u_{i-1})\frac{\lambda(m,u_{i})}{\lambda(u_{i})}
 \frac{\mathbb{1}(I(T)\geq n_{I}(T))}{\text{P}(I(T)\geq n_{I}(T)\mid \mathbf{H}(0))},
\end{align}
where $u_{0}=0$, $\lambda(u_{i})=(\alpha(u_{i})+\nu)I(u_{i-1})+\gamma E(u_{i-1})$ and $f(u_{i} \mid u_{i-1})=\lambda(u_{i})e^{-\int^{u_{i}}_{u_{i-1}}\lambda(u)du}$ is the conditional density evaluated at $u_{i}$.  Assuming $\alpha(u)$ is a continuous function, and $I(u),E(u)$ are right continuous, then according to Section 2.2, the jump (or mark) probabilities are:\\

\begin{align*}\frac{\lambda(m,u_{i})}{\lambda(u_{i})}=P(\mathbf{H}(u_{i})\mid \mathbf{H}(u_{i-1}))& = \begin{cases}
   \frac{\alpha(u_{i})I(u_{i-1})}{(\alpha(u_{i})+\nu)I(u_{i-1})+\gamma E(u_{i-1})} & \text{ if } m=1 \text{ (birth)}\\
   \frac{\gamma E(u_{i-1})}{(\alpha(u_{i})+\nu)I(u_{i-1})+\gamma E(u_{i-1})} & \text{ if } m=2 \text{ (migration)}\\
   \frac{\nu I(u_{i-1})}{(\alpha(u_{i})+\nu)I(u_{i-1})+\gamma E(u_{i-1})} & \text{ if } m=3 \text{ (death).}
\end{cases}
\end{align*}

The likelihood of the joint process is then:

\begin{align} \label{eq:lik1}
    \mathcal{L}^{*}(\lambda, \nu, \alpha; \mathbf{H}(u),\mathbf{C}(u))& =e^{-\int^{T}_{u_{k}}\lambda(u)du}\prod^{k}_{i=1}f(u_{i}\mid u_{i-1})\frac{\lambda(m,u_{i})}{\lambda(u_{i})}\text{P}(\mathbf{C}(u_{1}),\ldots,\mathbf{C}(u_{k})\mid \{\mathbf{H}(u_{i})\}^{k}_{i=1},\mathbf{C}(T))\times\\
    & \frac{\mathbb{1}(I(T)\geq n_{I}(T))}{\text{P}(I(T)\geq n_{I}(T)\mid \mathbf{H}(0))}, 
\end{align}

Re-arranging terms in Eq.~\ref{eq:lik1}, simplifying, and expressing the conditional joint distribution of $\{\mathbf{C}(u_{i})\}^{k}_{i=1}$ as a product of conditional probabilities, we get:

\begin{align} \label{eq:lik2}
    \mathcal{L}^{*}(\lambda, \nu, \alpha; \mathbf{H}(u),\mathbf{C}(u))& =e^{-\int^{u_{1}}_{0}\lambda(u)du}\prod^{1}_{i=k}\text{P}(\mathbf{C}(u_{i})\mid \mathbf{C}(u_{i+1}))\lambda(m,u_{i})
    e^{-\int^{u_{i+1}}_{u_{i}}\lambda(u)du} \frac{\mathbbm{1}(I(T)\geq n_{I}(T))}{\text{P}(I(T)\geq n_{I}(T)\mid \mathbf{H}(0))}\nonumber \\
    &=e^{-\int^{u_{1}}_{0}\lambda(u)du}\prod^{1}_{i=k}\mu(M,u_{i})
    e^{-\int^{u_{i+1}}_{u_{i}}\lambda(u)du}\frac{\mathbbm{1}(I(T)\geq n_{I}(T))}{\text{P}(I(T)\geq n_{I}(T)\mid \mathbf{H}(0))},
\end{align}

\noindent where $u_{k+1}=T$, and $\mu(M,u_{i})=P(\mathbf{C}(u_{i})\mid \mathbf{C}(u_{i+1}))\lambda(m,u_{i})$ with $M$ denoting a new mark that considers both the $m$ mark of the population EI process and the mark of the coalescent jump process (corresponding to transition probabilities of Eqs.~\ref{eq:trans1}-\ref{eq:trans5}). In particular, $M\in \{1,\ldots,6\}$ corresponds to 1: population birth and sample coalescent, 2: population birth and sample migration (E to I backwards transition), 3: birth that does not change coalescence state, 4: Migration in the population (E to I forwards) and in the sample (I to E backwards), 5: Migration in the population but not in the sample, and 6: population death. To be more precise, these marked rates are:

\begin{align*}
\mu(M,u_{i})& =\begin{cases}
   \alpha(u_{i})I(u_{i-1})\frac{n_{E}(u_{i+1})}{E(u_{i})}\frac{n_{I}(u_{i+1})}{I(u_{i})} & \text{ if } M(u_{i})=1 \text{ (birth and coalescent)}\\
  \alpha(u_{i})I(u_{i-1})\frac{n_{E}(u_{i+1})}{E(u_{i})}\frac{I(u_{i})-n_{I}(u_{i+1})}{I(u_{i})} & \text{ if } M(u_{i})=2 \text{ (birth and coalescent migration)}\\
   \alpha(u_{i})I(u_{i-1})\left(1-\frac{n_{E}(u_{i+1})}{E(u_{i})}\right) & \text{ if } M(u_{i})=3 \text{ (birth and off-sample event)}\\
   \gamma E(u_{i-1})\frac{n_{I}(u_{i+1})}{I(u_{i})} & \text{ if } M(u_{i})=4 \text{ (migration in the sample)}\\
   \gamma E(u_{i-1})\left(1-\frac{n_{I}(u_{i+1})}{I(u_{i})}\right) & \text{ if } M(u_{i})=5 \text{ (migration not in the sample)}\\
   \nu I(u_{i-1}) & \text{ if } M(u_{i})=6 \text{ (death).}
\end{cases}
\end{align*}
As we will show in the proof of Theorem 2.1 below, it is convenient to re-express $\mu(M,u_{i})$ in terms of the population sizes $I(u_{i})$ and $E(u_{i})$. Since knowing the type of the transition allows us to go from $I(u_{i-1})$ to $I(u_{i})$ (and similarly for $E(u_{i-1})$), we get:

\begin{align*}
\mu(M,u_{i})& =\begin{cases}
   \alpha(u_{i})I(u_{i})\frac{n_{E}(u_{i+1})}{E(u_{i})}\frac{n_{I}(u_{i+1})}{I(u_{i})} & \text{ if } M(u_{i})=1 \text{ (birth and coalescent)}\\
  \alpha(u_{i})I(u_{i})\frac{n_{E}(u_{i+1})}{E(u_{i})}\frac{I(u_{i})-n_{I}(u_{i+1})}{I(u_{i})} & \text{ if } M(u_{i})=2 \text{ (birth and coalescent migration)}\\
   \alpha(u_{i})I(u_{i})\left(1-\frac{n_{E}(u_{i+1})}{E(u_{i})}\right) & \text{ if } M(u_{i})=3 \text{ (birth and off-sample event)}\\
   \gamma (E(u_{i})+1)\frac{n_{I}(u_{i+1})}{I(u_{i})} & \text{ if } M(u_{i})=4 \text{ (migration in the sample)}\\
   \gamma (E(u_{i})+1)\left(1-\frac{n_{I}(u_{i+1})}{I(u_{i})}\right) & \text{ if } M(u_{i})=5 \text{ (migration not in the sample)}\\
   \nu (I(u_{i})+1) & \text{ if } M(u_{i})=6 \text{ (death).}
\end{cases}
\end{align*}
When there is a birth in the (forward) population process at time $u_{i}$, the corresponding rate is $\alpha(u_{i})I(u^{-}_{i})=\alpha(u_{i})I(u_{i-1})$ and the number of I lineages does not change at $u_{i}$, i.e. $I(u_{i})=I(u^{-}_{i})$ as only one E lineage is created. Similarly, when there is a migration E to I, the number of E lineages decreases by 1 and so $E(u^{-}_{i})=E(u_{i})+1.$

 Although the likelihood of Eq.~\ref{eq:lik2} has the same factorization of a time-reversal Markov process, this does not correspond to the time-reversal of the original EI population process with coalescent markings. A time-reversal expression would require calculation of the marginal of the population EI  process at present time $T$ and of transition probabilities as functions of multiple marginals.
We will now consider the scenario where the EI population sizes can be  approximated by the corresponding deterministic model governed by ordinary differential equations. Then, assumming we are given $\tilde{I}(u)$ and $\tilde{E}(u)$ (approximated at a pre-specified grid points in $[0,T]$), the likelihood corresponds to the likelihood of an Inhomogeneous Marked Poisson Process, that is:

\begin{align} \label{eq:lik3}
    \bar{\mathcal{L}}(\lambda, \nu, \alpha; k,\{u_{i},M(u_{i})\}^{k}_{i=1}
    )& =e^{-\int^{u_{1}}_{0}\tilde{\lambda}(u)du}\prod^{1}_{i=k}\tilde{\mu}(M,u_{i})
    e^{-\int^{u_{i+1}}_{u_{i}}\tilde{\lambda}(u)du}\mathbb{1}(\tilde{I}(T)\geq n_{I}(T)),
\end{align}
where $\tilde{\lambda}(u)$ and $\tilde{\mu}(M,u)$ are based on $\tilde{I}(u)$ and $\tilde{E}(u)$. 
We decouple the values of $I(u)$ and $E(u)$ from
    the jumps of the point process in order to propose an approximate
    backwards time point process.
Without this assumption, the backwards time process is a lazy discrete process (Eqs.\ref{eq:trans1}-\ref{eq:trans5}), a jump chain with possible jumps at times where $E(u)$ and $I(u)$ changed.    
Instead, we propose an approximate backwards time point process
    $\mathbf{\tilde{C}}(t)$, whose likelihood is derived from Equation
  \ref{eq:lik3}.

\begin{theorem} Let $\tilde{I}(t)$ and $\tilde{E}(t)$ approximate the EI population trajectories $I(t)$ and $E(t)$ on $[0,T]$, then conditioned on $\tilde{I}(t)$ and $\tilde{E}(t)$, the continuous time processes $\mathbf{\tilde{C}}(t)$ is described by the following  infinitesimal transition probabilities:
\begin{align*}
\text{P}(\mathbf{\tilde{C}}(t + dt) = (n_{E} - 1, n_{I}) | \mathbf{\tilde{C}}(t) = (n_{E}, n_{I}))&= n_{E} \times n_{I} \times \frac{\alpha(t)}{\tilde{E}(t)}dt + o(dt). \\
P(\mathbf{\tilde{C}}(t + dt) = (n_{E} + 1, n_{I} - 1) | \mathbf{\tilde{C}}(t) = (n_{E}, n_{I})) &= n_{I} \times \frac{\gamma (\tilde{E}(t)+1)}{\tilde{I}(t)} dt + o(dt) \\
P(\mathbf{\tilde{C}}(t + dt) = (n_{E} - 1, n_{I} + 1) | \mathbf{\tilde{C}}(t) = (n_{E}, n_{I})) &= \mathbbm{1}(\tilde{I}(t) > n_{I}(t)) \times n_{E} \times (\tilde{I}(t) - n_{I}) \times  \frac{\alpha(t)}{\tilde{E}(t)}dt + o(dt).
\end{align*}
Here $\mathbbm{1}(\tilde{I}(t)> n_{I}(t))$ is the indicator function which is 1 when $\tilde{I}(t) > n_{I}(t)$ and 0 otherwise. \end{theorem}\label{thm:coal}

\begin{proof}
We first change the direction of time (from present time 0 at the tips to the root) and express the likelihood of Eq.~\ref{eq:lik3} in terms of  $t_{i}=T-u_{k-i+1}$ for all $i=0,\ldots,k+1$
\begin{align} \label{eq:lik4}
    \mathcal{L}(\tilde{I},\tilde{E}, \nu, \alpha; k,\{t_{i},M(t_{i})\}^{k}_{i=1}
    )& =e^{-\int^{T}_{t_{k}}\tilde{\lambda}(t)dt}\prod^{k}_{i=1}\tilde{\mu}_{-}(M,t_{i})
    e^{-\int^{t_{i}}_{t_{i-1}}\tilde{\lambda}(t)dt}\mathbbm{1}(\tilde{I}(0)\geq n_{I}(0)),
\end{align}
where $\tilde{\mu}_{-}(M,t_{i})$  (and hence $\tilde{\lambda}(u)$) is expressed in terms of approximated population trajectories $\tilde{I}(t_{i})$, $\tilde{E}(t_{i})$, which are left continuous in reverse time (from present time 0 to the past $T$), and $\alpha(t_{i})$ evaluated at $t_{i}$, and coalescent lineage counts $n_{E}(t_{i-1})$, $n_{I}(t_{i-1})$. In particular, the background intensity function
\[\tilde{\lambda}(t)=(\alpha(t)+\nu)\tilde{I}(t)+\gamma \tilde{E}(t)+\gamma+\nu\]
is predictable \citep{daley2003introduction}.
Let $O=\{t^{*}_{1},t^{*}_{2},\ldots,t^{*}_{N}\} \subset \{t_{1},\ldots,t_{k}\}$ denote the ordered subset of event times with marks 1,2, and 4. Then the restricted point process to $O$ is the EI sample continuous-time coalescent process corresponding to the thinned Poisson process \citep{kingman1992poisson} with total rate 
\[\lambda_{c}(t)=\frac{\alpha(t)\tilde{I}(t)}{\tilde{E}(t)}n_{E}(t^{-})+\frac{\gamma(\tilde{E}(t)+1)}{\tilde{I}(t)}n_{I}(t^{-}),\]
and the corresponding marks have transition rates that correspond to the desired result.

\end{proof}
The likelihood of the $\tilde{\mathbf{C}}(t)$ process is then the corresponding likelihood of a continuous-time Markov chain with rates described in Theorem 2.1. In our implementation, we allow tips to have different sampling times (heterochronous sampling). Accommodating for heterochronous sampling is equivalent to allowing the $\mathbf{C}(t)$ chain to have deterministic jumps at sampling times. Finally, the corresponding likelihood for the EI sample genealogy is the $\tilde{\mathbf{C}}(t)$ likelihood divided by a combinatorial factor that counts the number of labeled (and colored) tree topologies. In the next sections, we only consider the likelihood of $\tilde{\mathbf{C}}(t)$.
\subsection{Augmented Likelihood}\label{sec:model_ll}

  Although our data is an unmarked genealogy where we only observe
  coalescent times (times of type 1 events) such as in Figure
  \ref{fig:Fig1}(C), our data-augmentation scheme allows us to sample
  the states $\mathbf{\tilde{C}}(t_{i})$ at a discrete grid of $N$
  times. We note that we do not augment the full realization of the
  $\mathbf{\tilde{C}}(t)$ process, instead we only sample the states
  at the grid times.
  \par
  For a partially observed $\mathbf{\tilde{C}}(t)$, the likelihood
  contribution of a coalescent interval can be written as a multiple
  risk phase type density (see \cite{hobolth2024phase} for an
    overview of phase type distributions for coalescent models, and
    \cite{lindqvist2023phase} for a general review of competing risk
  phase-type distributions).
  Under this framework, starting with $k$ lineages, the next
  coalescence is the first time the process reaches a state with only
  $k-1$ lineages, where each of the states with $k-1$ lineages are
  considered absorbing states.
  \par
  Define matrix $A_{k}$, a $k+1 \times k+1$ matrix which contains the
  rates of transitioning between states with $k$ lineages ordered so
  that the first row and column correspond to having $k$ lineages in
  state I, decreasing by 1 from left to right and top to bottom.
  In this case, there are $k+1$ total states, as there can be between
  $0$ and $k$ lineages in state I (vice versa state E).
  Let matrix $L_{k}$ be a $k+1 \times k-1$ matrix which contains the
  rates of transition to any of the $k-1$ possible coalescent states
  (any state with at least one lineage in each state can transition
  to coalescence, there are $k-1$ such states).
  Then the full intensity matrix $\mathbf{Q}_{k}$ describing the
  coalescent process from $k$ lineages to $k-1$ lineages is
  \begin{equation}
    \mathbf{Q}_{k} =
    \begin{bmatrix}
      \mathbf{A}_{k} & \mathbf{L}_{k} \\
      \mathbf{0} & \mathbf{0}
    \end{bmatrix}.
  \end{equation}
  In the case when the rates of $\mathbf{Q}_{k}$ are constant, the
  transition matrix is simply
  $\exp \mathbf{Q}_{k}t$, which by properties of block matrices, is
  \begin{equation}
    \exp \mathbf{Q}_{k}t =
    \begin{bmatrix}
      \exp \mathbf{A}_{k}t &  \mathbf{A^{-1}}_{k}(\exp
      \mathbf{A}_{k}t - \mathbf{I})\mathbf{L}_{k}\\
      \mathbf{0} & \mathbf{1}
    \end{bmatrix}.
  \end{equation}
  \par
  In pursuit of model simplicity, we will assume the rates are indeed
  constant between coalescent events.
  Let $s_{i}$ index the $k+1$ possible states of
  $\mathbf{\tilde{C}}(t_{i})$ with $k$ lineages.
  In particular, we define $s_{i}$ so that $s_{i} - 1$ is the number
  of lineages in state E (alternatively $k - s_{i} +1$ is the number
  of lineages in state I).
  Then the augmented likelihood contribution starting from time
  $t_{i-1}$ in state $s_{i-1}$ ending in coalescence at time $t_{i}$
  and state $s_{i}$ is
  \begin{equation}
    f_{s_{i-1}s_{i}}(t_{i} - t_{i-1}, x_{i} = 1) =
    \mathbf{e}_{s_{i-1}}^{T}\exp \{\mathbf{A}(t_{i} - t_{i-1})\}\mathbf{l}_{s_{i}}
  \end{equation}
  where $\mathbf{l}_{s_{i}}$ is the $s_{i}$th column of $\mathbf{L}$.
  In the case of heterochoronous sampling we also require augmented
  likelihood contributions for sampling events.
  These contributions are  calculated using the entries of $\exp
  \mathbf{A}(t_{i} - t_{i-1})$.
  Suppose at time $t_{i-1}$ the process is in state $s_{i-1}$ and the next time $t_{i}$ is a sampling event, then the process is in one
  of the $k+1$ transient states just before sampling, and after sampling the total number of lineages is increased deterministically by the number of samples collected at $t_{i}$.
  The likelihood contribution of such an event is
  $\exp{\mathbf{Q}(t_{i}-t_{i-1}})$ corresponding to the transition
  from $s_{i-1}$ to a transient state $s_{i}$ in the time $t_{i} - t_{i-1}$.
  We can write this contribution explicitly as
  \begin{equation*}
    f_{s_{i-1}s_{i}}(t_{i} - t_{i-1}, x_{i} = 0) =
    \exp{\mathbf{A}(t_{i}-t_{i-1})}_{s_{i-1},s_{i}}.
  \end{equation*}
  Let the states of the grid points be $\mathbf{s} = \{s_{0}, \dots, s_{N}\}$, then
  the augmented likelihood can be written as
  \begin{equation*}
P\left(\mathbf{t},\mathbf{x}, \mathbf{s}| \alpha, \gamma,
    \mathbf{\tilde{E}}, \mathbf{\tilde{I}}\right) = \prod_{i =
    1}^{N}f_{s_{i-1}s_{i}}(t_{i} - t_{i-1}, x_{i}).
  \end{equation*}
  \par
  The matrix exponential of $\mathbf{A}_{k}$ is not available in closed form.
  For full matrix exponentials, we use the scaling and squaring
  method of \cite{higham2005scaling} to numerically calculate the
  desired matrix.
  When only the action of the matrix exponential is needed for large
  matrices and small time steps, we use a Krylov subspace method
  \citep{niesen2009krylov}.
  We use the methods as implemented in the \texttt{julia} package
  \texttt{ExponentialUtilities.jl} \citep{DifferentialEquations.jl-2017}.

\subsection{Posterior Inference}
  \subsubsection{Coalescent EI Model Structure}
  We place a random walk prior on the grid of time-varying effective
  reproduction numbers $R_{u}$:
  \begin{align*}
    R_{0} &\sim \text{Log-Normal}(\mu_{0}, \sigma_{0}), \\
    \sigma &\sim \text{Log-Normal}(\mu_{rw},\sigma_{rw}), \\
    \log{(R_{k_{i}})}|R_{k_{i-1}},\sigma &\sim
    \text{Normal}(\log{(R_{k_{i-1}})}, \sigma).
  \end{align*}
  Let $\boldsymbol{\Theta} = (\gamma, \nu, I(0), E(0))$, $\mathbf{R}
  = (R_{0}, \dots, R_{k_{M}})$ be the vector of effective
  reproduction number values, $\mathbf{\tilde{E}}$ and
  $\mathbf{\tilde{I}}$ are the vectors of population counts.
  We augment the observed times $\mathbf{t}$ with additional
  pre-specified times which are the times at which the effective
  reproduction number is chosen to change.
  For this paper, we allow $R_{u}$ to change every seven days.
  In terms of the augmented likelihood, these additional grid times
  are treated as sampling events where the number of sampled lineages is 0.
  In the Appendix, we describe an alternative
  formulation of the model which augments the states only at the
  times of sampling and coalescence.
  Let $\mathbf{s}^{*} = (s_{0}, \dots, s_{N})$ be the states of
  $\mathbf{\tilde{C}}(t)$ at the pre-specified grid times $t^{*} =
  (t_{0}, \dots, t_{N})$ which includes $\mathbf{t}$.
  Let $\mathbf{x^{*}}$ be the corresponding vector of event types.
  The target posterior distribution is
  \begin{equation*}
    P(\textbf{R}, \textbf{s}^{*}, \boldsymbol{\Theta},
      \mathbf{\tilde{E}}, \mathbf{\tilde{I}}, \sigma \mid
    \textbf{x}^{*},\textbf{t}^{*}).
  \end{equation*}
  We estimate the target posterior via Metropolis within Gibbs
  sampling, by first sampling from
  \begin{align*}
    P(\textbf{R},\boldsymbol{\Theta}, \sigma, \mathbf{\tilde{E}},
      \mathbf{\tilde{I}} \mid
      \textbf{x}^{*},
      \textbf{t}^{*},
    \textbf{s}^{*}) &\propto P(    \textbf{x}^{*},
      \textbf{t}^{*},\textbf{s}^{*} \mid \mathbf{R},
    \boldsymbol{\Theta}, \mathbf{\tilde{E}},\mathbf{\tilde{I}})P(\mathbf{\tilde{E}},\mathbf{\tilde{I}}|\boldsymbol{\Theta}, \mathbf{R})
    P(\mathbf{R} \mid \sigma)P(\boldsymbol{\Theta},\sigma)
  \end{align*}
  via Markov chain Monte Carlo, then we sample from
  \begin{equation*}
    P(\textbf{s}^{*}|\textbf{R}, \boldsymbol{\Theta}, \sigma,
    \mathbf{\tilde{E}}, \mathbf{\tilde{I}}, \textbf{x}^{*}, \textbf{t}^{*})
  \end{equation*}
  via Monte Carlo by sequentially sampling the
  latent states directly as described in the following section.
  \par
  To sample the effective reproduction number and other population
  parameters via MCMC, we use elliptical slice sampling
  \citep{murray2010elliptical} with the augmented likelihood derived
  in the previous section, with a few modifications. After proposing a vector of parameters values, we obtain the values of $\tilde{E}(t)$ and $\tilde{I}(t)$ by solving the ordinary
  differential equation version of the EI population process
  described in Appendix Section \ref{sec:solveODE} and the next section. In our case, $P(\mathbf{\tilde{E}},\mathbf{\tilde{I}}|\boldsymbol{\Theta},
  \mathbf{R})$ is always equal to 1. The vector of parameter values and population size trajectories are then accepted jointly in the elliptical slice sampling algorithm. Further, we assign 0 likelihood to parameters which result in population
  counts exceeding 8 billion.
  We also assign 0 likelihood to parameters which result  in there
  being fewer members of the population than there are sampled
  lineages at the pre-specified times which contribute to the likelihood. The modified ESS algorithm is described in Algorithm 1 in
  the Appendix.
  
\subsubsection{ODE Approximation of the MJP}\label{sec:ODE}
  The system of equations which describes the deterministic approximation is
  \begin{align*}
    \frac{dE}{du} &= \alpha I - \gamma E, \\
    \frac{dI}{du} &= \gamma E - \nu I.
  \end{align*}
  In general, the system of ODEs can be thought of as the large
  population limit of the Markov Jump Process version of the EI
  population process \citep{kurtz1971limit,brittonandpardoux}.
  We solve this system of equations using the closed form solution
  derived in \cite{goldstein2025signal}, which in this case happens
  to be the solution to the conditional means of the Markov Jump Process.
  For our model, we assume initial conditions and allow $\alpha$ to be a piecewise constant
  time-varying parameter $\alpha(u)$ and then solve the ODE at the
  grid times $\mathbf{t}^{*}$.
  We assume E and I are piecewise constant between grid times.

  \subsubsection{Sampling latent states}\label{sec:condlatent}
  Define $S_{i}$ to be the random variable denoting the state of the
  process at the $i$th grid point at time $t_{i}$.
  The target conditional posterior distribution is
  \begin{equation*}
    P(S_{i} = s_{i}|S_{i-1}=s_{i-1}, T_{i-1} =t_{i-1}, T_{i}=t_{i},
    X_{i} = x_{i}, \alpha, \gamma, \mathbf{E}, \mathbf{I}).
  \end{equation*}
  Following the phase type framework described above,
  the joint density of $T_{i},S_{i}$ given $T_{i-1}, S_{i-1}, X_{i}, \alpha,
  \gamma, \mathbf{E}, \mathbf{I}$ is simply $f_{s_{i-1},s_{i}}(t_{i}
  - t_{i-1}, x_{i})$.
  By definition of conditional probability, then, we can calculate
  the conditional posterior as
  \small
  \begin{equation*}
    P(S_{i} = s_{i}|S_{i-1}, T_{i-1}, T_{i}, X_{i}, \alpha, \gamma,
    \mathbf{E}, \mathbf{I}) = \frac{ f_{s_{i-1}s_{i}}(t_{i} -
    t_{i-1}, x_{i})}{\sum_{j}f_{s_{i-1}s_{j}}(t_{i} - t_{i-1}, x_{i})}.
  \end{equation*}
  \normalsize

  Because we assume that individuals are only sampled in state I,
  $S_{0}$ is always known.
  This allows us to sample from the conditional posterior sequentially.

\subsection{Simulation Protocol}\label{sec:sim_proc}
    We simulate epidemics using a forward-time agent based stochastic
    SEIR model.
    An agent based SEIR model is an N-dimensional continuous time
    Markov chain, where N is the fixed population size.
    Each individual in the population can be in any of four states, S
    susceptible to infection, E infected but not yet infectious, I
    infectious, or R recovered/removed (no longer infectious).
    We use $\mathbf{G}(u)$ to denote the vector that records the
    states of all $N$ individuals in the population. If 
    $\mathbf{G}(u)_{i}$ is $S$, then the $ith$ individual is
    susceptible at time $u$.
    The process transitions from state $\mathbf{G}$ to state
    $\mathbf{G}'$ with rates:
    \begin{equation*}
      \lambda_{\mathbf{GG}'} =
      \begin{cases}
        \beta_{u}/N \times I(u) \text{ if $\mathbf{G}_{j} = S$ and
        $\mathbf{G}'_{j} = E$} \\
        \gamma \text{ if $\mathbf{G}_{j} = E$ and $\mathbf{G}'_{j} = I$} \\
        \nu \text{ if $\mathbf{G}_{j} = I$ and $\mathbf{G}'_{j} = R$
        }\\
        0 \text{ otherwise}
      \end{cases}
    \end{equation*}
    Here, $I(u)$ is the total number of individuals in state I at time $u$.
    We use the Gillespie algorithm \citep{gillespie1977exact} to
    simulate realizations of this model.
    When aggregated to a population level, this model is equivalent
    to the commonly used population-level SEIR model.
    \par
    We modify the algorithm to track infection histories, each time
    an individual becomes infected (changes from $S$ to E), we choose
    an individual in state I uniformly at random to be their
    infection source and record the infection history of the newly
    infected individual. For each individual, we also record the
    times they became infected, infectious, and recovered.
    \par
    For the population SEIR model, the  basic reproduction number,
    $R_{0}$, and effective reproduction number, $R_{u}$, are defined as
    \begin{align*}
      R_{0,u} = \frac{\beta_{u}}{\nu},  R_{u} =  R_{0,u} \times \frac{S(u)}{N},
    \end{align*}
    where $S(u)$ is the total number of individuals susceptible at time $u$.
    The basic reproduction number, $R_{0,u}$ is the average number of
    individuals an individual infected at time $u$ would subsequently
    infect in a completely susceptible population.
    The effective reproduction number $R_{u}$ is the average number
    of individuals an individual infected at time $u$ would
    subsequently infect if conditions remained the same as they were
    at time $u$.
    By setting $\alpha({u}) = \beta_{u}S(u)/N$, we are still able to
    calculate $R_{u}$ using the EI model.
    The parameter $1/\gamma$ is the average time spent infected but
    not infectious, it was set to 4 for all simulations.
    Likewise the parameter $1/\nu$ is the average time spent
    infectious, it was set to 7 for all simulations.
    The total population size was $N = 15000$, with one individual
    starting in state I and all other individuals starting in state $S$.
    \par
    Simulations were run for 22 weeks, and the last date of sampling
    was one day before 22 weeks.
    We simulated three separate $R_{u}$ trajectories, in the first
    setting (Fixed), $R_{0,u}$ was set at 2.2.
    This trajectory represents a situation where a new disease is
    introduced into a susceptible population and no behavior changes
    or health interventions occur.
    In the second trajectory (Increase), $R_{0,u}$ changed from 1.3
    to 2.3 over a five week period.
    This trajectory represents a situation where a disease becomes
    suddenly much more infectious, due to lifting of public health
    interventions, or changes in population behavior.
    In the final trajectory (Control), $R_{0,u}$ was set to 2.2 and
    then changed to 1.1 12 weeks into the simulation, representing a
    sudden public health intervention which rapidly reduces infectiousness.
    \par
    For each trajectory, we use two sampling schemes. In isochronous
    sampling (Iso), all samples are taken at the last date of sampling.
    In heterochronous (Het) sampling, some number of samples are
    sampled uniformly at random in the five week interval prior to
    the last date of sampling.
    Once an individual has been sampled, any subsequent individuals
    infected by the sampled individual (directly or indirectly) after
    the date of sampling cannot be sampled during subsequent sampling.
    For each type of sampling, we sampled either 50 or 100 total
    individuals from the currently infectious population completely at random.
    Starting from the last sampling time, a coalescent tree is formed
    deterministically using the infection histories of sampled individuals.
    The data given to the model are the coalescent and sampling
    times, as well as the number of lineages sampled at each time.
    For each trajectory, sampling scheme, and number of samples
    (twelve total scenarios), we simulated fifty trees.
    An example epidemic realization, $R_{u}$ curve, and genealogy are
    shown in Appendix Figure \ref{fig:example_sim_data}.
    Posterior distribution summaries were estimated using MCMC runs
    initially of length 100000 or 150000 with the first half
    discarded as burn in, thinned by every 10th/15th sample.
    Models were run until parameters had an ESS, tail ESS and bulk
    ESS of at least 100 \citep{vehtari2021rank}.
    As needed, the model was rerun with increased iterations and/or
    using a different seed.
    All code needed to reproduce the analyses in this paper are
    available at \url{https://github.com/igoldsteinh/ei_coal}.
\section{Results}
  \subsection{Fidelity of the EI Coalescent Approximation}
  To evaluate the fidelity of our derived EI coalescent model, we
  empirically compared the distributions of the true EI coalescent
  times and their distribution under our model. 
  We simulated 1000
  population trajectories from the individual (forward) EI population process
  (equivalent to the process described in section \ref{sec:sim_proc}
  for EI rather than SEIR models).
  Population processes were simulated with $\alpha = 2/3$, $\nu =
  1/3$, $\gamma = 1/2$, with one individual initially in the I
  compartment. Simulations were run till time 35, at which point they
  were rejected if there were less than 5 infectious individuals in
  the population. 
  If accepted, 5 infectious individuals were sampled
  uniformly at random, and their genealogy constructed
  deterministically using the known history of each individual (Empirical).
  Then, using the same population trajectories, an EI coalescent tree
  was simulated using the rate parameters of the model derived in
  Theorem~\ref{thm:coal} via time transformation  \citep{slatkin1991pairwise,hein2004gene}
  (TT).
  However, we rejected any simulations where the number of sampled lineages exceeded the corresponding population count, matching the model actually used for inference.
  Finally, a coalescent tree was simulated using the solution to the
  ODE rather than the true population trajectories (TT ODE).
  \par
\begin{figure}[ht]
    \includegraphics[width=\textwidth]{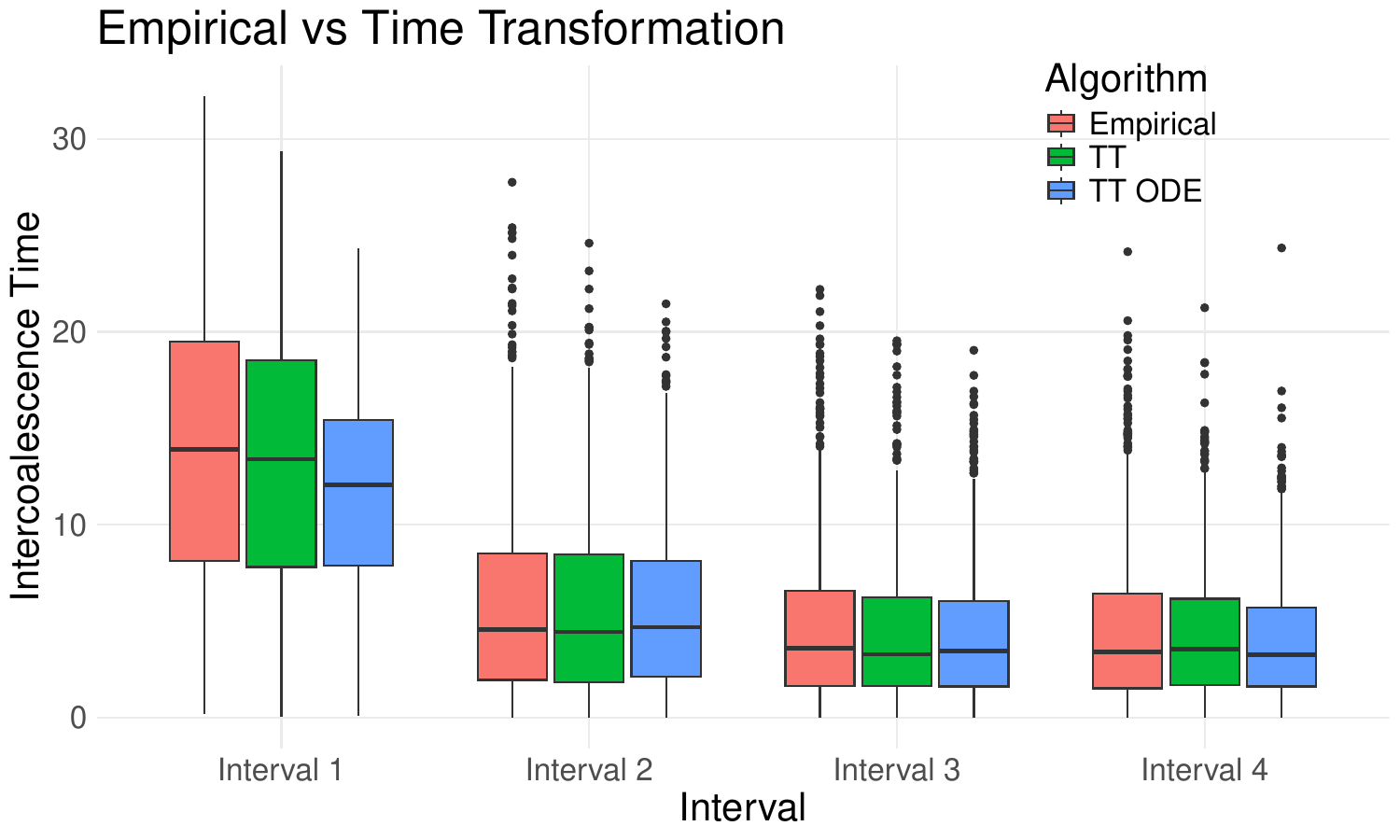}
    \caption{Distribution of intercoalescent intervals for 1000 true
      EI genealogies and 1000 approximate EI coalescent genealogies.
      Red (Empirical) boxes show the distribution for the true EI
      genealogies. Green (TT) boxes show the distribution for the
      observed EI coalescent genealogies generated via time transformation.
      Blue (TT ODE) boxes show the distribution for the partially
      observe coalescent genealogies using the ODE solution rather
      than the true stochastic trajectory.
      Genealogies are for five sampled individuals, thus there are 4
      total coalescent times, and 4 intercoalescent intervals.
      Intervals are ordered in backwards time, so that the end of
    interval 4 is the time to the most recent common ancestor of the genealogy.}
    \label{fig:empEI_vs_coalEI}
  \end{figure}

  The results of our simulation experiments are shown in Figure
  \ref{fig:empEI_vs_coalEI}. The distributions of the
  intercoalescence intervals under the true EI genealogies generally
  match that of the derived EI coalescent model when using the true
  population trajectories.
  However, when we instead use the ODE solution to simulate
  coalescent times, the distribution of coalescent times is more biased
  for some intervals.
  This is likely a result of the ODE solution not capturing the
  variation in the stochastic realizations of the EI population process.
  Appendix Figure \ref{fig:ode_vs_stoch} shows the ODE solution and ten of the 1000
  stochastic realizations previously described.
  At time 35, the ODE solution has approximately 250 individuals in
  both the E and I compartments, while the stochastic realizations
  range from over 1000 to less than 100 individuals in the E and I compartments.

  \subsection{Simulation Results}
  We simulated epidemics and coalescent genealogies using a
  forward-time agent based stochastic SEIR models which track
  individual infection histories. Using the simulated histories we then
  reconstructed true coalescent trees.
  We simulated under three different $R_{u}$ curves (Fixed, Increase
  and Control), two sampling schemes (Isochronous and Heterechronous)
  and two sample sizes (fifty and one hundred).
  There were fifty simulations for each of the twelve settings, full
  details are available in the methods section.
  Examples of the six simulations with fifty sampled lineages are
  visualized in Appendix Figure \ref{fig:samp50_plot}.
  The model's credible intervals successfully cover the true
  trajectories, although depending on the sampled tree, there can be
  wide uncertainty near the present day.
  We conjecture this uncertainty is a function of the number of
  coalescent events close to the present day.
  An example where the model can become both biased and more
  uncertain in the absence of coalescent events near the present day
  is shown in Appendix Figure \ref{fig:example_fail}.
  Examples of the six simulations with one hundred sampled lineages
  are visualized in Appendix Figure \ref{fig:samp100_plot}.
  \par
  We summarize model performance across fifty simulations for each
  scenario using frequentist metrics summarized in Table \ref{tab:sim_res}.
  \begin{table}[tbhp]
    \centering
    \small
    \begin{tabular}{rrrrrrrrrrrr}
      \hline
      Sim & ENV  & AD & MCIW  \\
      \hline
      Fix. Iso N=50 &  0.98 (0.66, 1.00) & 0.22 (0.17, 0.48) & 1.19
      (1.02, 1.92)  \\
      Fix. Iso N=100 &  0.96 (0.93, 0.98) & 0.18 (0.15, 0.26) & 0.96
      (0.90, 1.07)  \\
      Fix. Het N=50 & 0.93 (0.59, 1.00) & 0.27 (0.18, 0.80) & 1.34
      (1.10, 2.80) \\
      Fix.  Het N=100 & 0.96 (0.77, 0.98) & 0.20 (0.16, 0.27) & 1.02
      (0.96, 1.17)  \\
      Inc. Iso N=50 & 0.99 (0.73, 1.00) & 0.24 (0.14, 0.52) & 1.38
      (1.10, 2.20)  \\
      Inc. Iso N=100 & 0.99 (0.76, 1.00) & 0.17 (0.12, 0.30) & 1.09
      (0.96, 1.34) \\
      Inc. Het N=50 & 0.98 (0.75, 1.00) & 0.27 (0.16, 0.67) & 1.48
      (1.16, 2.59)  \\
      Inc. Het N=100 & 0.99 (0.75, 1.00) & 0.19 (0.13, 0.37) & 1.16
      (0.97, 1.44)  \\
      Ctrl Iso N=50 & 0.99 (0.92, 1.99) & 0.17 (0.12, 0.26) & 1.19
      (0.97, 1.44) \\
      Ctrl Iso N=100 & 0.98 (0.94, 1.00) & 0.16 (0.13, 0.22) & 0.94
      (0.89, 1.03)  \\
      Ctrl Het N=50 &  1.00 (0.85, 1.00) & 0.17 (0.11, 0.39) & 1.27
      (1.06, 1.83) \\
      Ctrl Het N=100 &  0.98 (0.90, 1.00) & 0.16 (0.13, 0.23) & 0.96
      (0.87, 1.05) \\
      \hline
    \end{tabular}
    \normalsize
    \caption{Summaries of freqentist metrics of model performance
      across fifty simulations. For the last three columns the first
      number is the medians, the numbers in parentheses are the 2.5\%
      and 97.5\% quantiles. Envelope (ENV) is a measure of coverage,
      and ideally should be 0.95. Absolute deviation (AD) is a measure
      of bias, smaller is better. Mean credible interval width (MCIW)
    summarises the widths of the 95\% credible intervals. }
    \label{tab:sim_res}
  \end{table}
  Envelope (ENV) is a measure of coverage, and is the proportion of
  time points for which an 95\% credible interval from the posterior
  distribution captured the true value of interest.
  Mean credible interval width (MCIW) is the mean of 95\% credible
  interval widths across time points within a simulation.
  Absolute deviation (AD) is a measure of bias, and is the mean of
  the absolute difference between the posterior median and the true
  value at each time point.
  We measure these quantities for each simulation using a grid of
  true $R_{u}$ values with a half-daily time-scale.
  Overall, we see that, as expected, with more sampled lineages, the
  model has lower deviation, narrower credible intervals, and
  reasonable, though often conservative coverage.
  With fifty lineages, the model still has generally good coverage,
  although performance can vary substantially for any particular simulation.

  \subsection{The Effective Reproduction number of Ebola in Liberia} 
Ebola disease is a severe illness caused by Ebola virus, which can spread from infected animals to humans, as well as from human to human transmission, resulting in regular outbreaks \citep{whoebola}. It is well known to have a significant latent period, plausibly between 6 and 11 days on average \citep{chowell2014transmission,velasquez2015time}. During the 2013 - 2016 Ebola epidemic in West Africa, more than 28000 cases and 11000 deaths were reported in the countries of Guinea, Liberia and Sierra Leone\citep{whowestafrica}.
In \cite{tang2023fitting}, the authors inferred 
  the time-varying basic reproduction number (Section
  \ref{sec:sim_proc}) from a pruned posterior tree of Ebola pathogen
  genomes constructed in \cite{dudas2017virus} using a coalescent
  model for the SIR epidemic model, which has no latent period and
  assumes individuals become infectious as soon as they are infected. 
  We conduct a similar analysis by estimating a maximum clade credibility tree of the two hundred and eight   Liberia sequences collected from the time period used in
  \cite{tang2023fitting} 
 (June 20th 2014 through February 14th 2015) collected in 
 \cite{dudas2017virus} and using that tree as input for our model. 
We estimate the maximum clade credibility tree using BEAST X
  \citep{baele2025beast} using the exact model specified in
  \cite{dudas2017virus}.
  In brief: the model used an HKY substitution model with four
  independent regions (codon positions one, two, three and non-coding
  intergenic regions), a relaxed molecular clock, and a
  non-parametric coalescent `Skygrid' model with one hundred change
  points \citep{gill2013improving}.
  Full details are available in \cite{dudas2017virus}.
  We assume this maximum clade credibility tree (MCC tree) is our data, and infer the effective reproduction number using our EI Coalescent model. 
  \par
\begin{figure}[ht]
    \includegraphics[width=\textwidth]{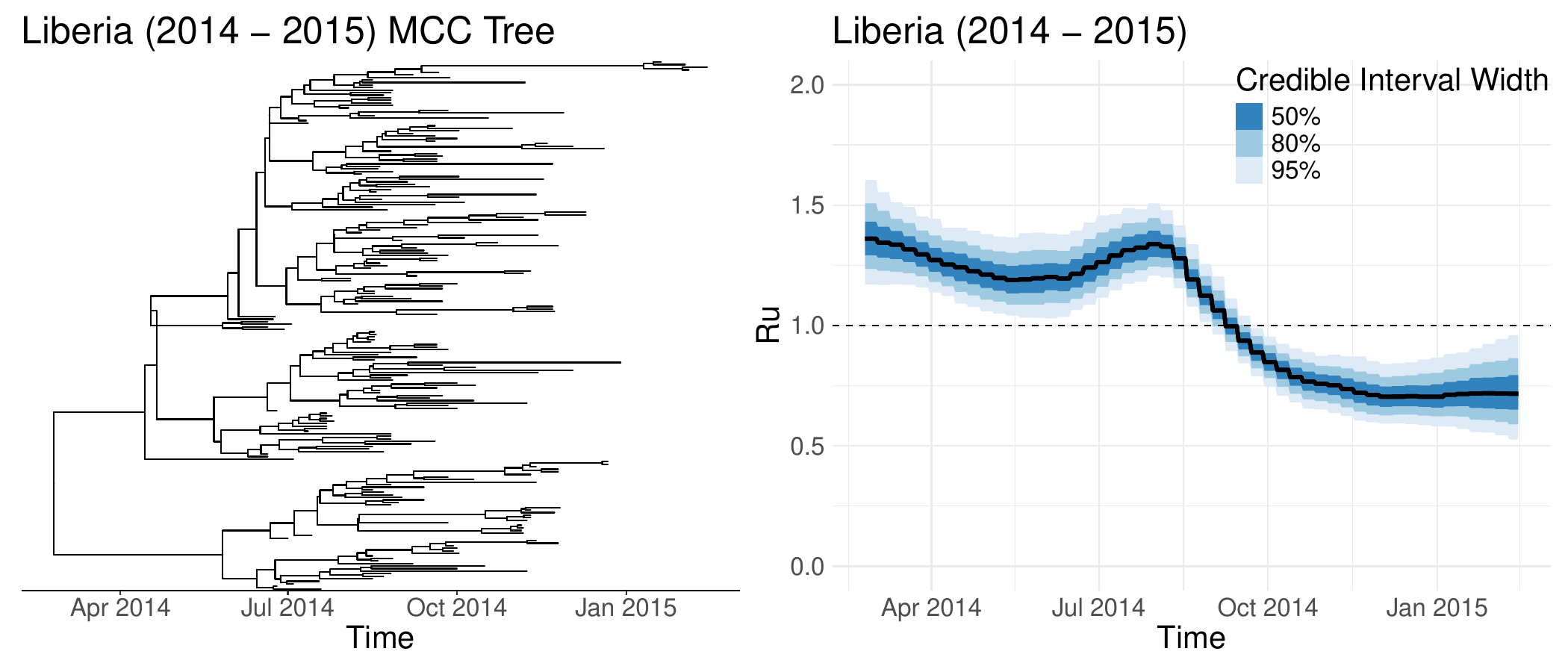}
    \caption{Maximum clade credibility tree and posterior summaries
      of the effective reproduction number for the 2014 Ebola epidemic
      in Liberia. In the right plot, black lines are posterior medians,
    blue shaded areas represent varying levels of credible intervals.}
    \label{fig:ei_liberia}
  \end{figure}

  For the EI Coalescent model, we used similar priors to those chosen
  in \cite{tang2023fitting}.
  The prior for $R_{u}$ was centered at 0.7, and the random walk
  standard deviation which was centered at 0.05.
  We chose to let $R_{u}$ change on a weekly basis, the
  \cite{tang2023fitting} analysis divided their time period into 40
  equal intervals of length 6.9 days.
  The prior for the infectious period was centered at seven days, and
  based on a prior analysis of Ebola in West Africa, we chose to also
  center the prior of the latent period at seven days \citep{fintzi2022linear}.
  The full details of prior specification used for the analysis are available in Appendix Table 2.
  Our maximum clade credibility tree (MCC tree) and posterior inference of the effective reproduction number are shown in Figure \ref{fig:ei_liberia}. 
  In the previous analysis, \cite{tang2023fitting} found the initial
  basic reproduction number was plausibly between 1.29 and 2.24, our
  upper bound for the initial effective reproduction number is
  substantially smaller, although overall we still find that the
  initial effective reproduction number is very likely to be above one.
  In addition to inferring different quantities, the two analyses are
  not directly comparable, as our MCC tree is not the same as the one
  used in \cite{tang2023fitting} (which was not available to directly
  analyze), and their analysis begins around May of 2014, while ours
  begins in February of 2014.
  Despite this difference, the broad trends of the two analyses are
  similar, as both our analyses agree $R_{u}$ was likely below one by
  October 2014.

\section{Discussion}
  In this paper we constructed the joint EI population and coalescent
  process, and an approximate model, the EI coalescent model,
  suitable for inference.
  We tested the performance of the inference model on simulations
  using fixed trees, demonstrating the model's general good
  performance.
  We applied our method to reanalyze the
  2014 Ebola outbreak in Liberia, obtaining parameter estimates within a
  plausible range of known behavior of Ebola and in agreement with previous analyses. 
Ebola is a good disease for our model, as it is well known to have
  a significant latent period \citep{velasquez2015time}.

  While previous work derived coalescent models
  for compartmental 
  deterministic systems \citep{volz2012complex}, and
  more recent work connects coalescent
  models to stochastic single-type birth-death processes \citep{crespo2021coalescent}, our work contributes in extending these previous efforts by considering a
  stochastic two-type birth-death process, the EI population process.
  Our construction of the joint process as a marked point process
  helps to clarify the nature of coalescent models for epidemic population processes and in particular, clarifies key differences to other approaches such as BDS processes. The techniques used to construct the joint EI population and
  coalescent process can be easily extended to other common epidemic
  models, such as the SEIR model.
  
  Although our experiments show that our proposed method has good empirical performance, we observed some discrepancy between the true coalescent distribution and the coalescent distribution using the ODE solution (Figure
  \ref{fig:empEI_vs_coalEI}). A stochastic model for the
  latent trajectories, such as a linear noise approximation
  \citep{tang2023fitting} might further improve model performance,
  albeit at the cost of increased computational burden.
  \par
  In general we observed good robustness performance, that is our model performed well in terms of frequentist
  metrics when fit to genealogies generated using a different data
  generating model. 
  This is a positive sign the model will perform well under
  non-trivial model misspecification.
  However, model performance can vary depending on the shape
  of the $R_{u}$ curve and the sampling scheme (Table \ref{tab:sim_res}).
  We hypothesize one driver of variable performance is how the
  coalescent events are distributed across time.
  If there are few coalescent events, the model may be unable to
  differentiate between a scenario where $E(t)$ and $I(t)$ are large,
  and a scenario when $\alpha({t})$ is small.
  \par
  The likelihood as written is evaluated
  via repeated matrix exponentiation, which is computationally
  expensive \citep{molerandvanloan}.
  At this time, we do not recommend using this model for trees with
  more than a few hundred tips.
  A future direction is to explore augmenting the entire EI coalescent process via uniformization so that the augmented likelihood does not involve matrix exponentiation. 
\par 
  Many of the parameters of the model, as with other compartmental models, are not identifiable, and we rely on informative priors to produce useful inference (Appendix \ref{fig:prior_post_plot}). 
  This is typical for this inference problem, for most models which infer the effective reproduction number, the generation time distribution, a function of the latent and infectious periods \citep{Svensson2007}, is fixed based on prior analyses, see for instance the popular EpiNow2 package \citep{epinow2}. 
  Data integration may help make more model parameters identifiable, since our model conditions on the latent states of the epidemic process, it could be easily adapted to conduct joint inference from a genealogy and other typical data sources such as counts of cases or hospitalizations.
  
  Finally, our method assumes a given genealogy is available without error. This is an unrealistic situation since genealogies are usually not directly observed. In future research, we plan to incorporate our methodology within BEAST to estimate model parameters and population trajectories from molecular sequences directly.

\section{Acknowledgments}
I.H.G was supported by a Stanford Center for Computational, Evolutionary and Human Genomics Fellowship.  J.A.P. acknowledges support from the NSF Career Award \#2143242 and NIH Award R35GM148338.
Thanks to Nicola Muller for helpful discussion. 
\bibliographystyle{abbrvnat}
\bibliography{references}  
\clearpage

\appendix

\setcounter{table}{0}
\setcounter{equation}{0}
\setcounter{section}{0}
\setcounter{figure}{0}

\renewcommand\thefigure{\thesection\-\arabic{figure}}
\renewcommand\thetable{\thesection\-\arabic{table}}

\section{Appendix}
\subsection{Compartmental Models}\label{app:comp_models}
\subsubsection{The SEIR Model} \label {seir_model}
The SEIR model is the traditional model used for a pathogen with a latent period. 
In this section, we describe it in detail, and show its relationship to the EI model used in the main paper. 
The SEIR describes an infectious disease outbreak of a homogeneously mixing population, with the population divided into four compartments: susceptible, exposed, infectious, and removed.
We represent the SEIR model as a four dimensional continuous time Markov jump process, $\mathbf{G(u)} = (S(u), E(u), I(u), R(u))$.
By construction, $R(u)$ is redundant, as $R(u) = N - S(u) - E(u) - I(u)$, where $N$ is the fixed total population size.
The SEIR dynamics are described by rate parameters such that 
\begin{align*}
    P(\mathbf{G}(u + du) &= (s - 1, e + 1, i, r)  \mid \mathbf{G}(u) = (s,e,i,r)) = \beta_{u} i s/N du + o(du),\\
    P(\mathbf{G}(u + du) &= (s, e - 1, i + 1, r) \mid \mathbf{G}(u) = (s,e,i,r)) =  \gamma e  du + o(du), \\ 
    P(\mathbf{G}(u + du) &= (s, e, i - 1, r + 1) \mid \mathbf{G}(u) = (s,e,i,r)) = \nu  i du + o(du),\\
    P(\mathbf{G}(u + du) &= (s, e, i, r ) \mid \mathbf{G}(u) = (s,e,i,r)) = 1 - (\beta i  s/N  + \gamma  e  + \nu  i)du + o(du). 
\end{align*}
Here $\gamma$ is the inverse of the mean latent period, and $\nu$ is the inverse of the mean infectious period. 
We describe the infectiousness of the disease through the time-varying rate parameter $\beta_{u}$.
With this model, the time-varying basic reproduction number, $R_{0,u}$, and effective reproduction number, $R_{u}$, are defined as
\begin{align*}
R_{0,u} = \frac{\beta_{u}}{\nu},  \quad R_{u} =  R_{0,u} \times \frac{S(u)}{N}.
\end{align*}
The basic reproduction number, $R_{0,u}$ is the expected number of individuals an individual infected at time $u$ would subsequently infect in a completely susceptible population. 
The effective reproduction number $R_{u}$ is the expected number of individuals an individual infected at time $u$ would subsequently infect if conditions remained the same as they were at time $u$. 
Intuitively, $R_{u}$ takes into account the fact that some people have already been infected, as it is $R_{0,u}$ multiplied by the fraction of the population which is still susceptible at time $u$.
\subsubsection{The EI Model}
The epidemic model used in the main text is the EI model, which is represented as a two dimensional continuous time Markov jump process $\mathbf{H}(u) = (E(u), I(u))$, defined as:
\begin{align*}
    P(\mathbf{H}(u + du) &= (e + 1, i)  \mid \mathbf{H}(u) = (e,i)) = \alpha_{u} i  du + o(du),\\
    P(\mathbf{H}(u + du) &= (e - 1, i + 1) \mid \mathbf{H}(u) = (e,i)) =  \gamma  e du + o(du), \\ 
    P(\mathbf{H}(u + du) &= (e, i - 1) \mid \mathbf{H}(u) = (e,i)) = \nu  i  du + o(du),\\
    P(\mathbf{H}(t + du) &= (e, i) \mid \mathbf{H}(u) = (e,i)) = 1- (\alpha_{u}  u + \gamma  u  + \nu \times i) du + o(du).
\end{align*}
In contrast to the SEIR model, there is no S compartment, rather, changes in the susceptibility of the population are incorporated in changes in $\alpha_{u}$.
Likewise, there is no R compartment, however, the dynamics of recovery remain the same as in the SEIR model, we just don't track the cumulative number of recoveries.
The effective reproduction number, $R_{u}$ is still recoverable by setting $\alpha_{u} = \beta_{u} \times \frac{S(u)}{N}$, so that
\begin{align*}
R_{u} =  R_{0,u} \times \frac{S(u)}{N}  = \frac{\alpha_{u}}{\nu}.
\end{align*}
\subsection{Closed Form Solution to the EI ODE System}\label{sec:solveODE}
In \citep{goldstein2025signal}, we derived the closed form solution to the EI ODE system for a time interval where $\alpha$ is constant shown below using \texttt{Mathematica} (Version 13.1).
\begin{align*}
\frac{dE}{du} &= \alpha I - \gamma E, \\
\frac{dI}{du} &= \gamma E - \nu I.
\end{align*}
In particular, we treat the solution to the system as the action of the matrix exponential of 
\begin{equation*}
\mathbf{V}u = \begin{bmatrix}
-\gamma u & \alpha u \\
 \gamma u & -\nu u \\
\end{bmatrix}
\end{equation*}
on the vector $[E(0), I(0)]$. 
We then solve this action of the matrix exponential symbolically.
We provide the solutions below. 
Let $B = \sqrt{4\alpha\gamma + (\gamma - \nu)^{2}}$. 
Define the following four functions:
\begin{align*}
g_{1}(\gamma, \nu, \alpha, u) &= \frac{4\alpha(1+e^{B})\gamma + (\gamma - \nu)(1+e^{B}\gamma + B - \nu - e^{B}(B + \nu)}{2e^{(\gamma + B + \nu)u/2}(4\alpha\gamma + (\gamma - \nu)^{2})},\\
g_{2}(\gamma, \nu, \alpha, u) &= \frac{\alpha (-1+e^{B})}{e^{(\gamma + B + \nu)u/2}B}, \\
h_{1}(\gamma, \nu, \alpha, u) &= \frac{\gamma(-1 + e^{B})}{e^{(\gamma + B + \nu)u/2}B},\\
h_{2} &= \frac{4\alpha(1+e^{B}\gamma + (\gamma - \nu)*(\gamma - B + e^{B}B+e^{B}(\gamma-\nu)-\nu)}{2e^{(\gamma + B+ \nu)u/2}(4\alpha\gamma + (\gamma-\nu)^{2})}.
\end{align*}
Then, given the values of $E(0)$ and $I(0)$, we have
\begin{align*}
E(u) &= g_{1}(\gamma, \nu, \alpha, u)E(0) + g_{2}(\gamma, \nu, \alpha, u)I(0), \\
I(u) &= h_{1}(\gamma, \nu, \alpha, u)E(0) + h_{2}(\gamma, \nu, \alpha, u)I(0).
\end{align*}
\subsection{Rate Matrix Details for the EI Coalescent Likelihood}
Below we include an example $\mathbf{A}$ and $\mathbf{L}$ matrix for $n= 3$ extant lineages for clarity. 
Again the columns and rows are ordered so that the first column (row) corresponds to the state of all $n$ lineages in state I, decreasing by one from left to right and top to bottom. 
\small
\begin{equation*}
\mathbf{A} = \begin{bmatrix}
-3 \frac{\gamma (e+1)}{(i)} &  3 \frac{\gamma (e+1)}{(i)} & 0 & 0 \\
 (i - 2)^{+} \frac{\alpha}{(e)} & -\left((i - 2)^{+} \frac{\alpha}{e}) +2 \frac{\gamma (e+1)}{(i)}  + 2 \frac{\alpha}{e}\right) & 2 \frac{\gamma e}{(i)} & 0 & 0 \\
 0 & 2(i-1)^{+} \frac{\alpha}{(e)} & -\left(2(i-1)^{+} \frac{\alpha}{(e)} +  \frac{\gamma (e+1)}{(i)} + 2\frac{\alpha}{(e)} \right) & 2 \frac{\gamma e}{(i)}& 0 \\
 0 & 0 & 3i\frac{\alpha}{(e)} & -3i\frac{\alpha}{(e)}
\end{bmatrix}.
\end{equation*}
\begin{equation*}
\mathbf{L} = \begin{bmatrix}
0 & 0 \\
2 \frac{\alpha}{e} & 0 \\
0 & 2 \frac{\alpha}{e} \\
0 & 0 
\end{bmatrix}.
\end{equation*}
\subsection{Alternative EI Augmented Likelihood with Piecewise Time-varying Parameters}\label{sec:alt_ll}
In the main paper, we use a model which augments the observed data with the counts of lineages in the E and I compartments at coalescent, sampling, and pre-specified grid times at which $\alpha(t)$ is \textit{a priori} allowed to change. 
In this way, the likelihood is the product of terms corresponding to number of $\alpha(t)$ changepoints. 
Alternatively, one can write the augmented likelihood only 
at coalescent and sampling times, but not at the times when $\alpha(t)$ is assumed to change. 
In this case, suppose that for the interval that starts at $t_{i-1}$ in state $s_{i-1}$ and ends at time $t_{i}$ in state $s_{i}$ that there are $J$ times in the interval during which $\alpha$ changes. 
We denote the times at which $\alpha$ changes as $t_{i,1}, \dots t_{i, J-1}$ and $A_{k,j}$ is the matrix of rates for the $jth$ subinterval, likewise $L_{k,j}$ is the corresponding $L$ matrix. 
Then, by properties of Continuous Time Markov Chains, we can write the likelihood contributions for when $x_{i} = 1$ as follows:
\begin{equation*}
f_{s_{i-1},s_{i}}(t_{i-1}, t_{i,1}, \dots, t_{i,J}, t_{i}, x_{i} = 1) = \mathbf{e}^{T}_{s_{i-1}}\exp\mathbf{A}_{k,1}(t_{i,1} - t_{i-1})\exp\mathbf{A_{k,2}}(t_{i,2} - t_{i,1})\dots\exp\mathbf{A_{k,J}}(t_{i} - t_{i,J-1})\mathbf{l_{J, s_{i}}}.
\end{equation*}
Likewise, when $x_{i} = 0$, 
\begin{equation*}
f_{s_{i-1},s_{i}}(t_{i-1}, t_{i,1}, \dots, t_{i,J}, t_{i}, x_{i} = 0) = \exp\mathbf{A}_{k,1}(t_{i,1} - t_{i_-1})\exp\mathbf{A_{k,2}}(t_{i,2} - t_{i,1})\dots\exp\mathbf{A_{k,J}}(t_{i} - t_{i,J-1})_{s_{i-1},s_{i}}.
\end{equation*}
Using this alternate likelihood, less states need to be sampled (augmented), which may improve the efficiency of the MCMC algorithm. 
To test this empirically, we fit both approaches
to the simulated data used in Figure 2 and ran them for 30000 iterations on the same 2022 Macbook M2 Air. 
The minimum ESS per second for the alternative version of the model was 0.044, versus 0.043 for the main version of the model, suggesting any benefits are not dramatic. 
Both models are available on the github repo. 

\subsection{Elliptical Slice Sampling}
Let $\textbf{q}$ be the vector of accepted log scale mean zero parameters, that is $\textbf{q}$ is the vector of accepted $\boldsymbol{\Theta}$, $\bf{R}$, and $\sigma$ on the log scale with the prior mean (on the log scale) subtracted from their values.
All priors are log-normal, so on the log scale they are normal, and we can write a joint prior normal density with variance matrix $\Sigma$.
Let $\mathbf{\mu}$ be those prior means on the log scale. 
Let $\mathbf{\tilde{E}}, \mathbf{\tilde{I}}$ be the values of $\tilde{E}(t)$ and $\tilde{I}(t)$ needed to evaluate the likelihood for the accepted values $\textbf{q}$.
Define $L(\mathbf{q}, \mathbf{\tilde{E}}, \mathbf{\tilde{I}}) = \log{P(\mathbf{t}, \mathbf{x}, \mathbf{t}^{*}, \mathbf{s}|\exp{q + \mu}, \mathbf{\tilde{E}}, \mathbf{\tilde{I}})}$.
Further define $V(\mathbf{\tilde{E}}, \mathbf{\tilde{I}}, \mathbf{t}, \mathbf{x}, \mathbf{t}^*, \mathbf{s})$ be a function which is 1 if, for all points $t_{i}$ in $\mathbf{t}$ and $\mathbf{t*}$, $n_{E}(t_{i}) \leq \tilde{E}(t_{i})$  and $n_{I}(t_{i}) \leq \tilde{I}(t_{i})$ and 0 otherwise.
Finally, let $W(\mathbf{\tilde{E}}, \mathbf{\tilde{I}})$ be a function which is one if $\max{\mathbf{\tilde{E}} + \mathbf{\tilde{I}}} \leq 8E9$. 
The modified elliptical slice sampling algorithm \citep{murray2010elliptical} is as follows:
\begin{algorithm}[H]
\caption{Modified Elliptical Slice Sampling Algorithm with ODE Solution}\label{alg:ess}
 \hspace*{\algorithmicindent} \textbf{Input} Current state $\mathbf{q}$ prior means $\boldsymbol{\mu}$, compartment counts $\mathbf{\tilde{E}}, \mathbf{\tilde{I}}$\\
 \hspace*{\algorithmicindent} \textbf{Output} 
 New state $\mathbf{q'}$  updated $\mathbf{\tilde{E}}, \mathbf{\tilde{I}}$
\begin{algorithmic}[1]
\State \text{Sample ellipse:} $\boldsymbol{\psi} \sim N(0, \Sigma)$
\State \text{Set log likelihood threshold:}
\State $u \sim \text{Uniform}(0,1)$
\State $\log y \gets L(\mathbf{q}, \mathbf{\tilde{E}}, \mathbf{\tilde{I}}) + \log{u}$
\State \text{Draw the initial angle}
\State $\theta \sim \text{Uniform}(0, 2\pi)$
\State $[\theta_{min}, \theta_{max}] \gets [\theta - 2\pi, \theta]$
\State $\mathbf{q}' \gets \mathbf{q}\sin{\theta} + \boldsymbol{\psi}\cos{\theta}$
\State Update $\mathbf{\tilde{E}},\mathbf{\tilde{I}}$ using $\exp{\mathbf{q}'} + \boldsymbol{\mu}$ \Comment{ODE solution according to Section \ref{sec:solveODE}}
\If{$L(\mathbf{q}', \mathbf{\tilde{E}}, \mathbf{\tilde{I}}) > \log y$ \& $V(\mathbf{\tilde{E}}, \mathbf{\tilde{I}}, \mathbf{t}, \mathbf{x}, \mathbf{t}^*, \mathbf{s}) == 1$ \& $W(\mathbf{\tilde{E}}, \mathbf{\tilde{I}})$ == 1}
    \State return: $\mathbf{q'}, \mathbf{\tilde{E}}, \mathbf{\tilde{I}}$
\Else 
    \State Shrink the bracket:
    \If{$\theta < 0$} 
        \State $\theta_{min} \gets \theta$ 
    \Else
        \State $\theta_{max} \gets \theta$
    \EndIf
    \State $\theta \sim \text{Uniform}(\theta_{min}, \theta_{max})$
    \State \text{Go To 8}
\EndIf
\end{algorithmic}
\end{algorithm}

\section{Results}
\begin{figure}[ht] 
\begin{center}
\includegraphics[width=0.7\textwidth]{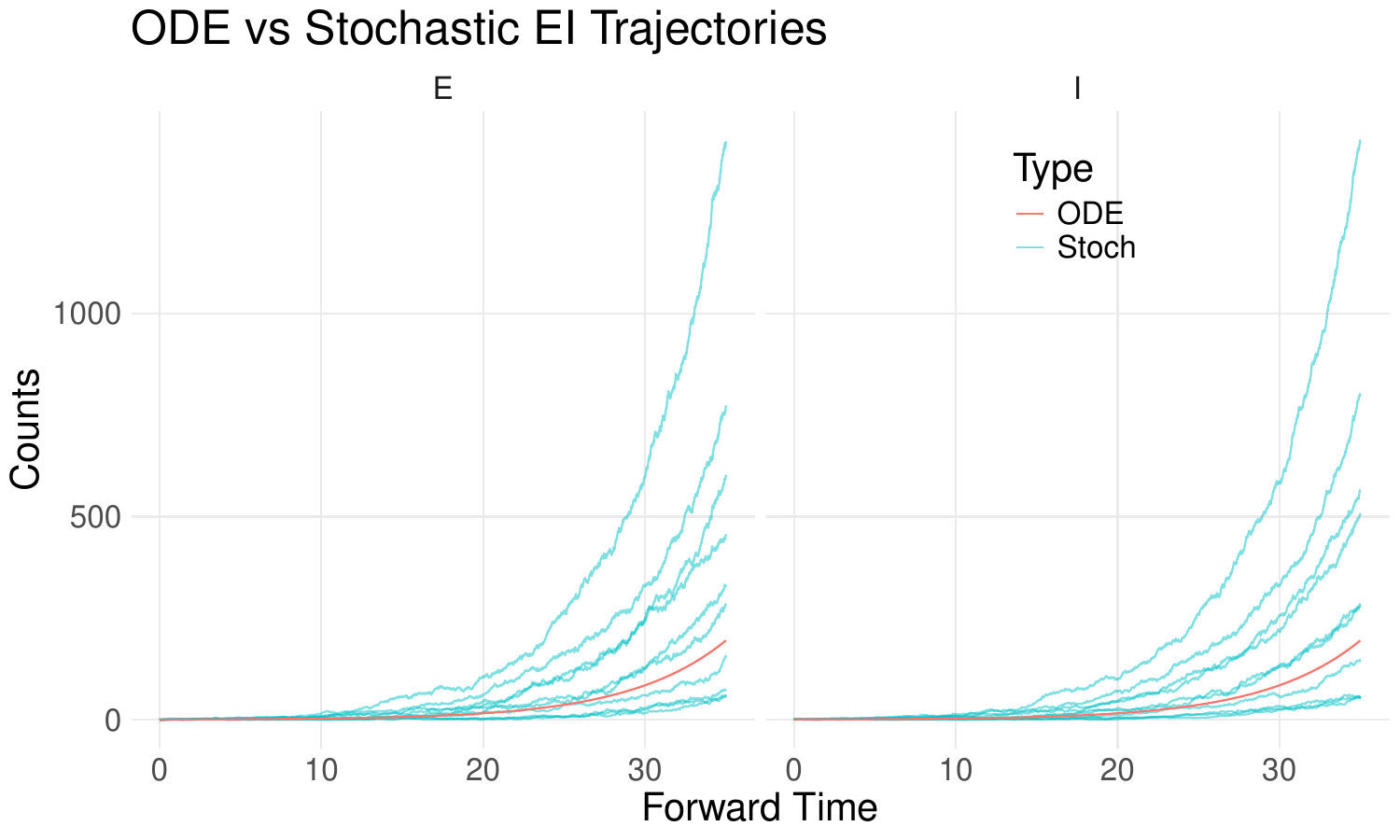}
  \caption{Example with 10 realizations of the stochastic EI population process (blue) and the ODE solution to the deterministic EI process (red).} \label{fig:ode_vs_stoch}
  \end{center}
\end{figure}

\subsection{Fidelity of the EI Coalescent Approximation}

\begin{figure}[H]
\includegraphics[width=1.0\textwidth]{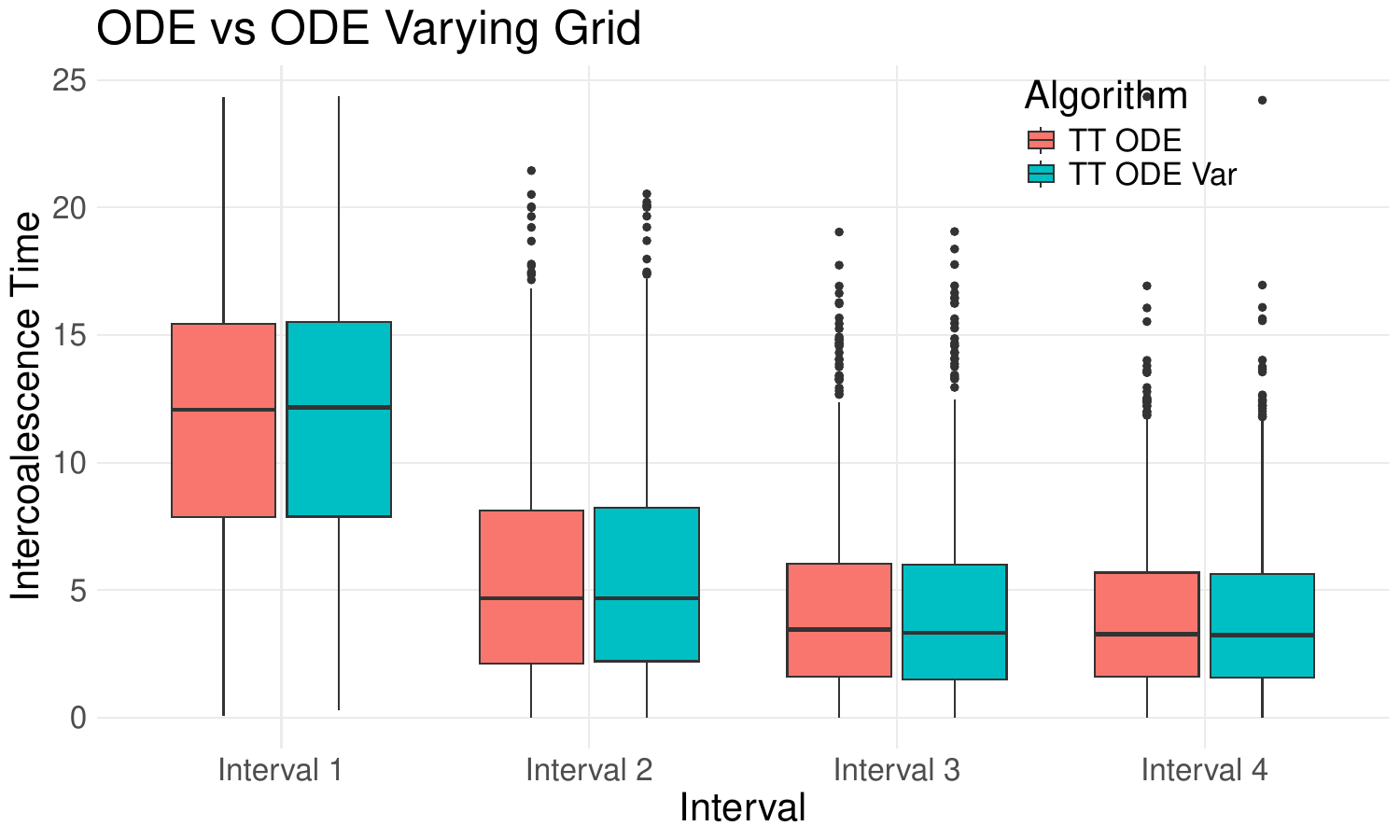}
  \caption{Boxplots of intercoalescent intervals. 
  Red (ODE) boxeplots correspond to  intercoalescent intervals simulated using an ODE solution solved on a regular grid (0.5, 1, 1.5, \dots). Blue (TT ODE Var) boxeplots correspond to intercoalescent intervals simulated using the ODE solution solved at the realized times simulated from the 1000 stochastic realizations of the EI process used to construct intervals in Figure 2 of the main text.
  Phylogenies with five tips are reconstructed, thus there are 4 intercoalescent intervals. 
  Intervals are ordered in backwards time, so that the end of interval 4 is the time to the most recent common ancestor of the phylogeny}
  \label{fig:ode_vs_var_ode}
\end{figure}

In Figure 2 we showed a comparison of the empirical distribution of coalescent times simulated from the true population dynamics and the empirical distribution of coalescent times simulated according to our proposed EI coalescent model. We argue that the discrepancy is mainly due to the ODE approximation to the population dynamics. In Figure~\ref{fig:ode_vs_stoch}, we show ten stochastic realizations of the EI population process and the ODE solution, providing a range of possible discrepancies between the two models.
In contrast, if we change the grid times at which we solve the ODE, the distribution of coalescent intervals remains largely unchanged. 
We tested this by comparing the distribution of coalescent intervals when simulating using the ODE solution solved at regular grid times (0.5, 1, 1.5, \dots) vs the realized change times simulated from the 1000 stochastic realizations of the EI model (Figure \ref{fig:ode_vs_var_ode}).

\subsection{Simulation Results}
\begin{figure}[ht]
\includegraphics[width=1.0\textwidth]{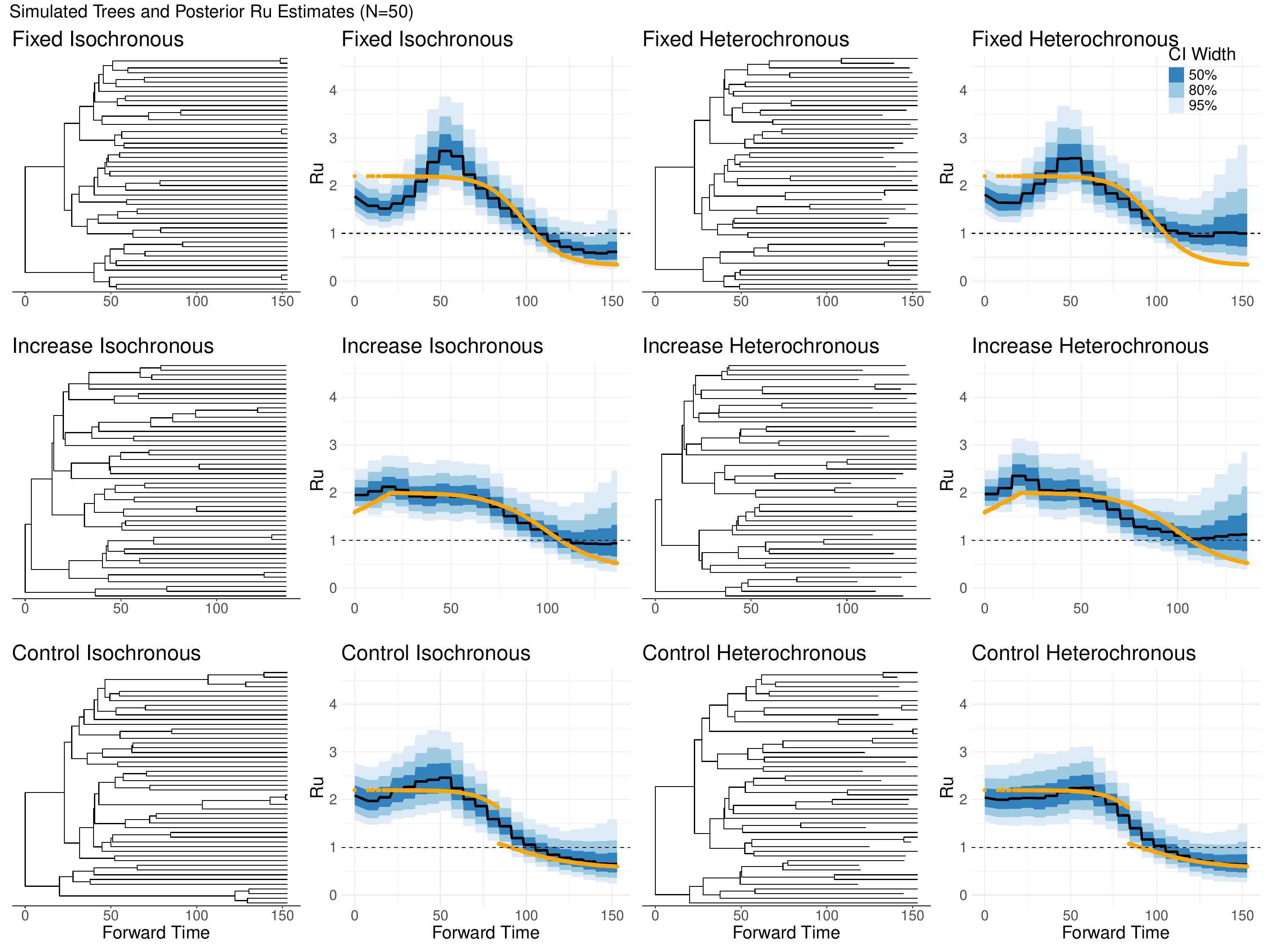}
  \caption{Simulated sampled genealogies and posterior summaries of $R_{u}$ for simulation scenarios with fifty sampled lineages.
  First row is the fixed scenario, second is the increase scenario, third row is the control scenario. True $R_{u}$ values are shown in orange, black lines are posterior medians, blue shaded areas are credible intervals.
  Time is in forward time, with time 0 being set to the time of the most recent common ancestor.
  Note the increase scenario has a different time axis, because the time to the most recent common ancestor is consistently smaller than in the other two scenarios.}
  \label{fig:samp50_plot}
\end{figure}

\begin{figure}[H]
\includegraphics[width=1.0\textwidth]{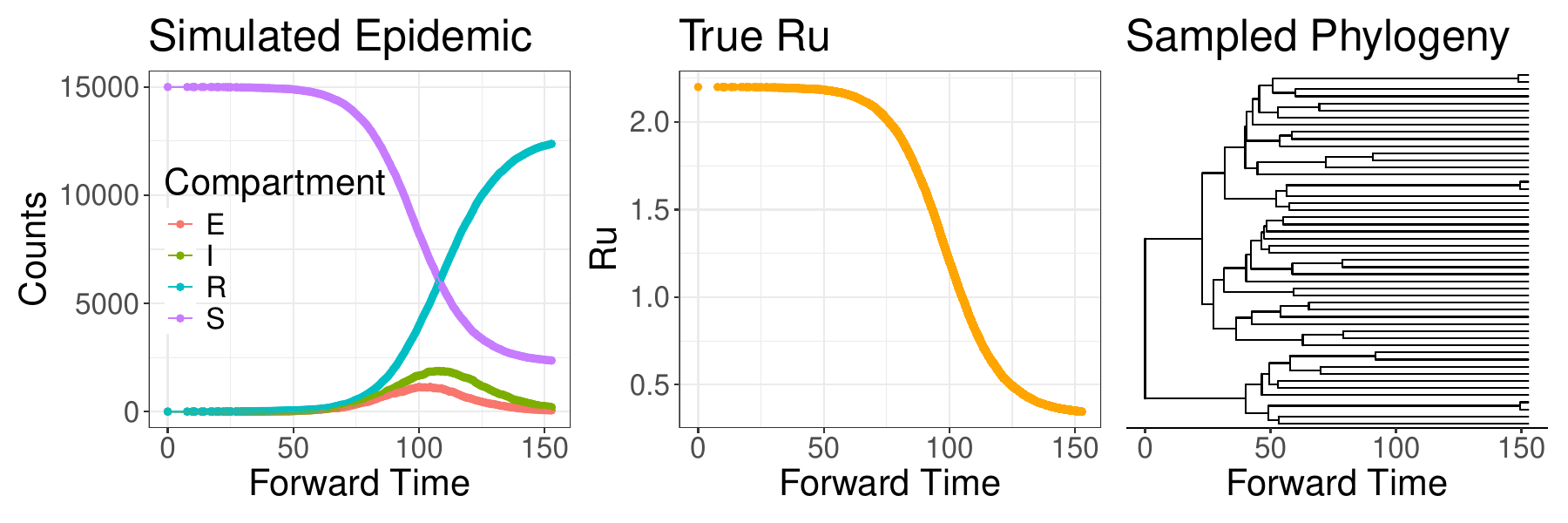}
  \caption{Example of a simulated epidemic, true $R_{u}$ curve, and sample genealogy. 
  The left-most plot shows the simulated Epidemic, different colors correspond to the counts of individuals in each state on a half-day time grid. 
  The middle panel shows the true $R_{u}$ curve which is the target of inference.
  The right panel shows the sample genealogy of 50  lineages.
  The genealogy was used to infer the posterior estimates of $R_{u}$ shown in the top row of the second column of Figure \ref{fig:samp50_plot}.}
  \label{fig:example_sim_data}
\end{figure}

\begin{figure}[H]
\includegraphics[width=1.0\textwidth]{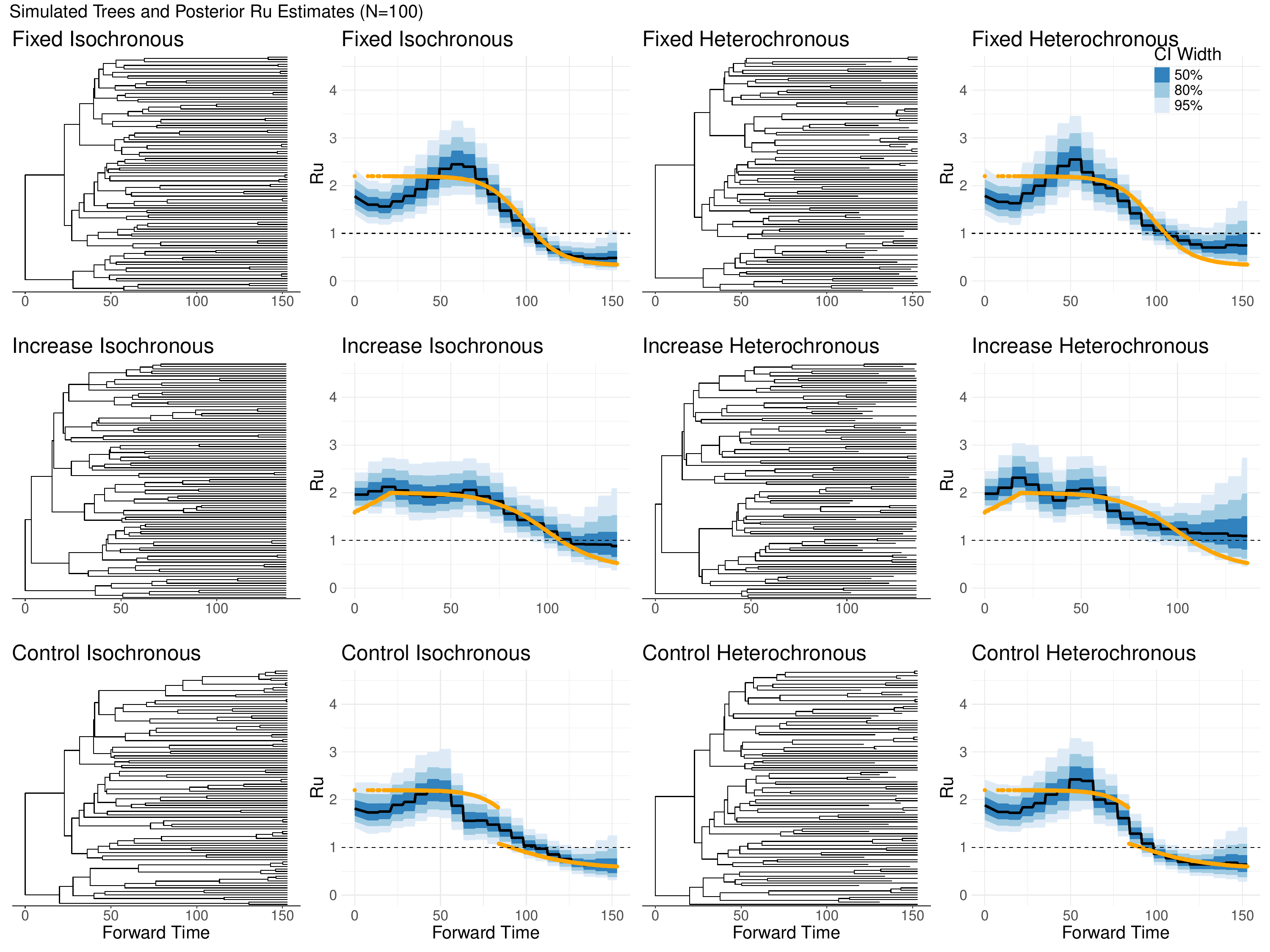}
  \caption{Simulated sampled genealogies and posterior summaries of $R_{u}$ for simulation scenarios with one hundred sampled lineages.
  First row is the fixed scenario, second is the increase scenario, third row is the control scenario. True $R_{u}$ values are shown in orange, black lines are posterior medians, blue shaded areas are credible intervals.
  Time is in forward time, with time 0 being set to the time of the most recent common ancestor.
  Note the increase scenario has a different time axis, because the time to the most recent common ancestor is consistently smaller than in the other two scenarios.}
  \label{fig:samp100_plot}
\end{figure}

Examples of the six simulations with fifty sampled lineages are visualized in Figure \ref{fig:samp50_plot}.
The model's credible intervals successfully cover the true trajectories, although depending on the sampled tree, there can be wide uncertainty near the present day. 
Figure \ref{fig:example_sim_data} shows an example simulated epidemic and sample genealogy. 
Figure \ref{fig:samp100_plot} shows examples of the six simulations with 100 sampled lineages. 


\subsubsection{Inference Failure}
We hypothesize this kind of failure can occur when there are not enough coalescences occuring in a particular region of the genealogy, in the case shown in Appendix Figure \ref{fig:example_fail} there is only one coalescence in the last 30 days of the simulation.  

\begin{figure}[H]
\includegraphics[width=1.0\textwidth]{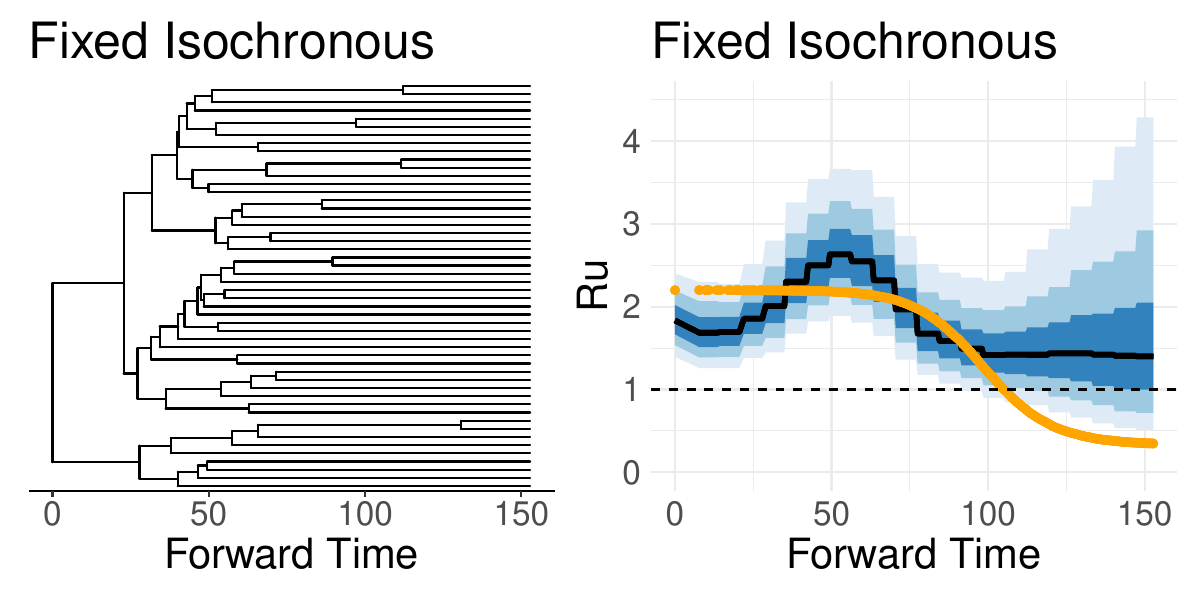}
  \caption{Example of a simulated sampled genealogy from the Fixed Iso simulation where the model fails to cover the true $R_{u}$ curve.
  Yellow dots are true $R_{u}$ values, black lines are posterior medians, blue shaded areas are varying levels of credible intervals (see \ref{fig:samp50_plot}).}
  \label{fig:example_fail}
\end{figure}

\subsubsection{Model Priors for Simulations}

\begin{figure}[H]
\begin{center}
\includegraphics[width=0.9\textwidth]{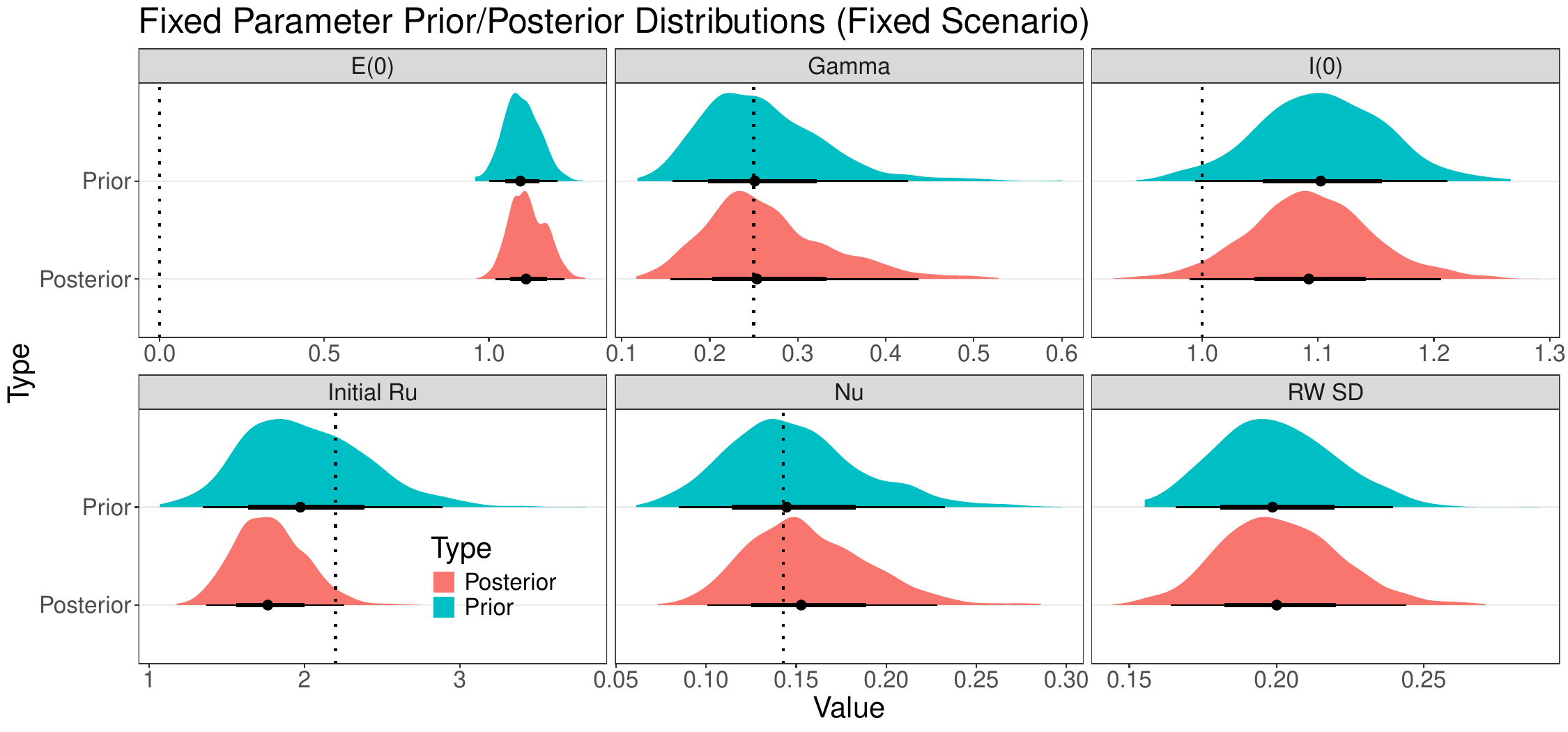}
  \caption{Prior and posterior densities for fixed model parameters. Posterior summaries are from the model fit to the data shown in Figure \ref{fig:samp50_plot} top left plot of the main text. 
    Blue densities are the prior, red densities are the posterior, dotted lines indicate true values.}
  \label{fig:prior_post_plot}
  \end{center}
\end{figure}

For the Increase scenario, the prior for $R_{0}$ was centered at 1.2 rather than 2, otherwise the priors were the same as the Fixed scenario. 
The priors for the Control scenario were the same as the Fixed scenario. 
Because the time to the most recent common ancestor is random, there is no fixed true value of the initial $R_{u}$ or for the initial counts in the E and I compartments across simulations. 
\
\begin{table}[H]
\caption{Priors used in the Fixed simulation scenarios.}
\centering
\small
\fbox{%
\begin{tabular}{*{5}{c}}
          Parameter & Model & Prior & Prior Median (95\% Interval) & Truth \\
         \hline \\
         $\gamma$ & All & Log-normal($\log(1/4)$, 0.25) & 0.25 (0.15, 0.4) & 0.25 \\
         $\nu$ &  All &  Log-normal($\log(1/7)$, 0.25) & 0.14 (0.09, 0.23) & 0.14  \\
         $\sigma_{rw}$ & All & Log-normal(log(0.2), 0.1) & 0.2 (0.16, 0.24) & NA \\
         $R_{0}$ &  All & Log-Normal(log(2.0), 0.2) & 2.0 (1.35, 2.99) & Varies \\
         $E(0)$ & All & Log-Normal(log(1.1), 0.05) & 1.1 (0.99, 1.21) & Varies \\
         $I(0)$ & All & Log-Normal(log(1.1), 0.05) & 1.1 (0.99, 1.21) & Varies \\
\end{tabular}}
\label{tab:model_sim_priors}
\end{table}
In Figure \ref{fig:prior_post_plot} we compare the prior and posteriors of the fixed parameters of the model. 
The initial number of latent and infectious individuals, as well as the average latent period and average infectious period are all not identifiable.
This is typical for this inference problem unfortunately, for most models which infer the effective reproduction number, the generation time distribution, a function of the latent and infectious periods \citep{Svensson2007}, is fixed based on prior analyses, see for instance the popular EpiNow2 package \citep{epinow2}. 
The prior for the random walk standard deviation must likewise be chosen by the user.

\subsubsection{Model Priors for Liberia Ebola analysis}
For the Ebola analysis, we used priors similar to those used in the analyses of \cite{tang2023fitting} and \cite{fintzi2022linear}.

\begin{table}[H]
\caption{Priors for the 2014 Liberia Ebola outbreak.}
\centering
\small
\fbox{
\begin{tabular}{*{5}{c}}
          Parameter & Model & Prior & Prior Median (95\% Interval) \\
         \hline \\
         $\gamma$ & All & Log-normal($\log(1/7)$, 0.45) & 0.14 (0.06, 0.35) \\
         $\nu$ &  All &  Log-normal($\log(1/7)$, 0.3) & 0.14 (0.08, 0.26)  \\
         $\sigma_{rw}$ & All & Log-normal(log(0.05), 0.2) & 0.05 (0.03, 0.07) \\
         $R_{0}$ &  All & Log-Normal(log(0.7), 0.5) & 0.7 (0.26, 1.87)  \\
         $E(0)$ & All & Log-Normal(log(1.1), 0.05) & 1.1 (0.99, 1.21) \\
         $I(0)$ & All & Log-Normal(log(1.1), 0.05) & 1.1 (0.99, 1.21) \\
\end{tabular}}
\label{tab:model_ebola_priors}
\end{table}

\end{document}